\documentclass[aps,pra,twocolumn,superscriptaddress,amsmath,amssymb,showpacs]{revtex4-1}

\usepackage{graphicx}	% Include figure files
\usepackage{dcolumn}	% Align table columns on decimal point
\usepackage{bm}		% Bold math
\usepackage{verbatim} 	% Long comments/Users/mpant/Dropbox (MIT)Qnetworkflow/Qnet.tex
\usepackage{gensymb}
\usepackage{braket}
\usepackage{xr}
\usepackage{hyperref}
\usepackage{enumerate}
\usepackage{color}
\usepackage{paralist}

\def\mymathhyphen{{\hbox{-}}}
\begin{document}

%\title{Enhanced Quantum Communication Rates with multi-path routing}
\title{Routing entanglement in the quantum internet}

%\pacs{42.50.Ex, 03.67.Dd, 03.67.Lx, 42.50.Dv}

\author{Mihir Pant}
\email{mpant@mit.edu}
\affiliation{Department of Electrical Engineering and Computer Science, MIT, Cambridge, MA}
\affiliation{Quantum Information Processing group, Raytheon BBN Technologies, 10 Moulton Street, Cambridge, MA}
\author{Hari Krovi}
\affiliation{Quantum Information Processing group, Raytheon BBN Technologies, 10 Moulton Street, Cambridge, MA}
\author{Don Towsley}
\affiliation{College of Information and Computer Sciences, University of Massachusetts, Amherst, MA}
\author{Leandros Tassiulas}
\affiliation{School of Engineering and Applied Science, Yale University, 17 Hill House Avenue, New Haven, CT}
\author{Liang Jiang}
\affiliation{Departments of Applied Physics and Physics, Yale University, New Haven, CT}
\affiliation{Yale Quantum Institute, Yale University, New Haven, CT}
\author{Prithwish Basu}
\affiliation{Advanced Networking Systems, Raytheon BBN Technologies, 10 Moulton Street, Cambridge, MA}
\author{Dirk Englund}
\affiliation{Department of Electrical Engineering and Computer Science, MIT, Cambridge, MA}
\author{Saikat Guha}
\affiliation{Quantum Information Processing group, Raytheon BBN Technologies, 10 Moulton Street, Cambridge, MA}
\affiliation{College of Optical Sciences, University of Arizona, 1630 East University Boulevard, Tucson, AZ}

\begin{abstract}
Remote quantum entanglement can enable numerous applications including distributed quantum computation, secure communication, and precision sensing. In this paper, we consider how a quantum network---nodes equipped with limited quantum processing capabilities connected via lossy optical links---can distribute high-rate entanglement simultaneously between multiple pairs of users (multiple flows). We develop protocols for such quantum ``repeater" nodes, which enable a pair of users to achieve large gains in entanglement rates over using a linear chain of quantum repeaters, by exploiting the diversity of multiple paths in the network. Additionally, we develop repeater protocols that enable multiple user pairs to generate entanglement simultaneously at rates that can far exceed what is possible with repeaters time sharing among assisting individual entanglement flows. Our results suggest that the early-stage development of quantum memories with short coherence times and implementations of probabilistic Bell-state measurements can have a much more profound impact on quantum networks than may be apparent from analyzing linear repeater chains. This framework should spur the development of a general quantum network theory, bringing together quantum memory physics, quantum information theory, and computer network theory.

\end{abstract}

\maketitle

A {\em quantum network} can generate, distribute and process quantum information in addition to classical data~\cite{2008.Nature.Kimble.QuantumInternet}. The most important function of a quantum network is to generate long distance quantum entanglement, which serves a number of tasks including the generation of multiparty shared secrets whose security relies only on the laws of physics~\cite{1991.PRL.Ekert.QCrypt, 2014.TheoCompSc.Bennett-Brassard.BB84}, distributed quantum computing~\cite{1999.PRA.Cirac-Macchiavello.DistQC}, improved sensing~\cite{2012.PRL.Gottesman-Croke.QRepTelescope, 2014.NatPhys.Komar-Lukin.QuantClocks}, blind quantum computing~(quantum computing on encrypted data)~\cite{2009.FOCS.Broadbent-Kashefi.BlindQC}, and secure private-bid auctions~\cite{2008.IJQI.Guha-Beausoleil.QAuction}. 

\begin{figure}[h]
\includegraphics[width=\columnwidth]{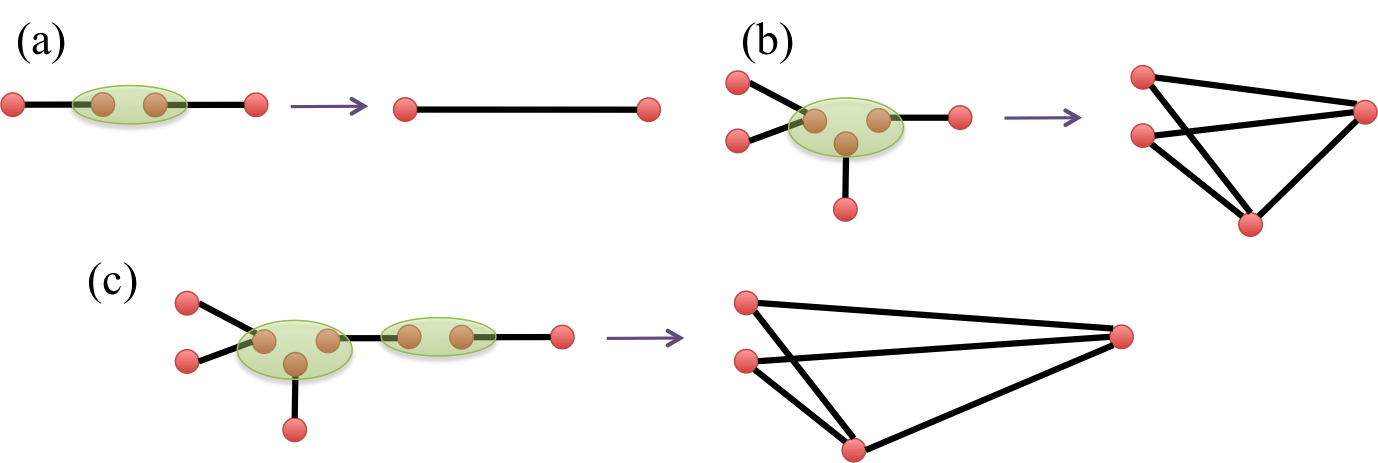}
\caption{Examples of fusing small entangled clusters into larger ones using projective quantum measurements (green ovals) at nodes of a quantum network. Red circles represent qubits and black lines represent entanglement. (a) Two-qubit measurement (Bell state projection) used to connect two entangled links into a longer entangled link; (b) a three-qubit measurement (a GHZ state projection) fuses three clusters (two $2$-qubit entangled links and one $3$-qubit linear cluster) into one $4$-qubit entangled cluster; (c) two adjacent nodes in a network performing a three-qubit GHZ measurement and a two-qubit Bell state measurement simultaneously. The measured qubits are lost, whereas the final entangled state of the unmeasured qubits upon successful completion of both measurements is the same regardless of the order or the simultaneity of the measurements. A quantum measurement at a node may succeed only with a probability, which is a function of the class of optical devices employed to realize the measurement (e.g., linear optics, single photon sources, and single photon detectors) and losses in devices. This figure does not show ``failure outcomes", i.e., the resulting entangled state if one or both measurements fail.}
\centering
\label{fig:fusionexamples}
\end{figure}

Recent experiments have demonstrated {\em entanglement links}, viz., long-range entanglement established between quantum memories separated by a few kilometers using a point-to-point optical link~\cite{2015.Nature.Hensen-Hanson.BellTest}. As illustrated in Fig.~\ref{fig:fusionexamples}, measurements performed at nodes in a quantum network can be used to glue together small entanglement links into longer-distance clusters. The nodes contain quantum memories that store qubits up to their coherence time, sources that generate photons entangled with the quantum memory to be sent to neighboring nodes, and local quantum processors that can perform multi-qubit joint measurements. Entanglement attempts between neighboring nodes are synchronized on a global clock. The {\em quantum routing protocol} dictates the measurements to be performed locally at each node in order to obtain the desired entanglement topology. Possible goals of a routing protocol could be to enable high rate entanglement among multiple user-pairs simultaneously, or to generate multi-partite entanglement (entanglement between three or more parties). 

The development of network algorithms and protocols for routing and scheduling information flows was critical for the creation of today's Internet. We expect a similar development in algorithm/protocol design to be critical to design a versatile and high performance quantum network. Some results and intuitions from the theory of classical networks carry over into quantum networking. However, many new challenges arise due to the idiosyncrasies of quantum information. Unlike classical communications, where the rate can be increased by increasing transmit power, photon loss fundamentally limits the entanglement rate over any single link, which must decay exponentially with the length of optical fiber, regardless of the choice of quantum source, the transmit power or the detection strategy~\cite{2014.NatureComm.Takeoka-Wilde.TGWBound, 2017.NatComm.Pirandola-Banchi.QuantCommUltRate}. Whereas copying of bits at a network node is common in multipath routing in classical networks~\cite{2000.TechRep.Vahdat-Becker.EpRoutAdhoc, 2005.Sigcomm.Biswas-Morrix.ExOR}, copying a qubit is impossible because of the quantum no-cloning theorem~\cite{1982.Nature.Wootters-Zurek.NoCloning, 1982.PLA.Dieks.EPRComm}. Unlike classical information flow, an entanglement flow does not have directionality. Rather, entanglement is generated across links all over the network and pieced together to form long-range entanglement. Quantum memories are much shorter lived and expensive compared to their classical counterparts making classical routing strategies such as disruption tolerant routing~\cite{2004.SigComm.Jain-Patra.DTNRout, 2006.InfoComm.Burgess-Levine.MaxProp}---where a packet is held by a node for until the desired next-hop link is up---much harder to mimic. Finally, distilling and shaping entanglement among a desired set of nodes from many copies of large (potentially random) entangled clusters is a purely quantum problem that has no classical analogue.

\begin{figure*}[t]
\includegraphics[width=0.7\textwidth]{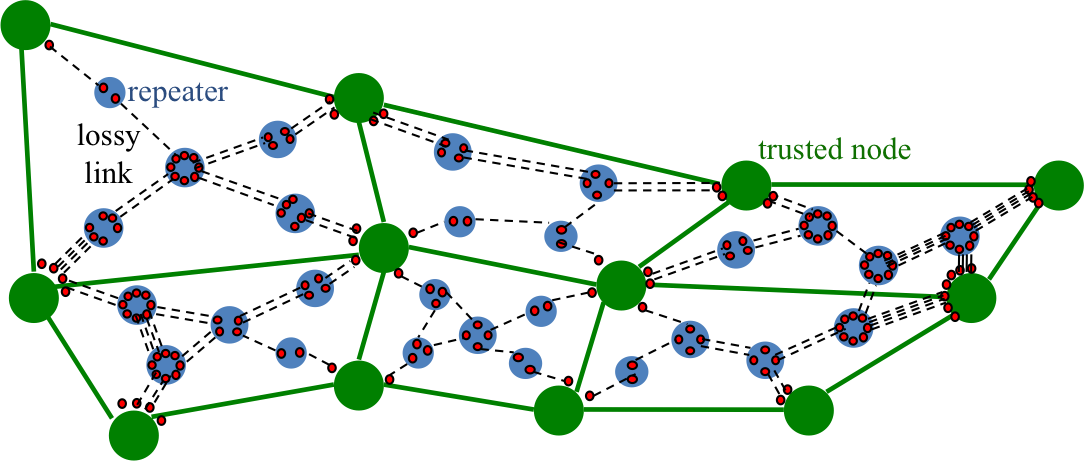}
\caption{Schematic of a general quantum repeater network. The large (green) circles represent `trusted' nodes, which are connected via a classical network. The blue circles denote repeater stations, and the red circles inside them represent quantum memories. The dashed lines connecting the red circles are independent lossy optical (fiber) channels. In principle, all nodes in the network could be equipped with quantum repeaters (i.e., no trusted nodes), in which case depending upon the need, a node can be a consumer of shared entanglement, or act a router to conduit entanglement flows between other nodes.}
\centering
\label{fig:USnetwork}
\end{figure*}

In this paper, we present protocols for repeater nodes to support multiple simultaneous entanglement flows when every node is limited to the same quantum processing used in repeater chains: (probabilistic) two-qubit Bell state measurement (BSM), also called entanglement swapping. BSMs have been experimentally demonstrated in many physical systems~\cite{2008.Nature.Yuan-Pan.BDCZExp, 2013.Nature.Bernien-Hanson.3ment, 2009.Science.Olmschenk-Monroe.QuantTelMatQubit, 1998.PRL.Pan-Zeilinger.EntSwapPhot, 2005.Nature.Chou-Kimble.MeasIndEnt, 2007.Nature.EntSingleAtomBits}. Entanglement attempts between repeater nodes are probabilistic because they are connected via lossy optical links. In every clock cycle, pairs of neighboring repeater nodes attempt entanglement generation. The result of whether an entanglement generation was successful is transmitted back to the corresponding pair of nodes. The repeater nodes then make their local BSM decisions based on this `local' link-state knowledge, i.e., the successes and failures of entanglements established across their nearest neighbor links. Measurements at different nodes can all be done in parallel because BSMs commute with one another.

One of our interesting findings is that {\em{multi-path routing}}, i.e. using multiple paths for routing entanglement in a quantum network, can enable long distance entanglement generation with a superior rate-vs.-distance {\em scaling} than a single linear repeater chain along the shortest path connecting Alice and Bob~\footnote{Pirandola recently showed~\cite{2016.ArXiv.Pirandola.CapRepQC}, for an information-theoretic description of repeaters that are ideal fully-error-corrected universal quantum processors, that the optimal rate attainable for multi-path entanglement routing using such ideal repeaters is superior to the rate of a linear chain of ideal repeaters.}. The rate along a single repeater chain falls off exponentially with distance~\cite{2015.PRA.Guha-Tittel.QRRateLossAnalysis, 2017.PRA.Pant-Guha.AllOptRepRescources}. {{Multi-path routing}} reduces the exponent, resulting in an exponential increase in rate with increasing distance. The quantum network we propose uses the same basic elements and operations (probabilistic BSMs) as a linear repeater chain but uses more repeaters. Note that increasing the number of repeaters in a repeater chain would not always increase the rate: given the total end-to-end distance, and given the losses at each node, there is an optimal number of repeaters between the end points of a flow that maximizes the rate, i.e., inserting more nodes along that linear path can actually diminish the rate~\cite{2015.PRA.Guha-Tittel.QRRateLossAnalysis, 2017.PRA.Pant-Guha.AllOptRepRescources}. 

%We would like to note here that Pirandola recently showed~\cite{2016.ArXiv.Pirandola.CapRepQC}, for an information-theoretic description of repeaters that are ideal fully-error-corrected universal quantum processors, that the optimal rate attainable for multi-path entanglement routing using such ideal repeaters is superior to the rate of a linear chain of ideal repeaters. 

Another interesting result is that if the repeater nodes have `global' link-state knowledge (knowledge of the state of all links in the network) and the entanglement generation probability is above a (percolation) threshold, multi-path routing enables long distance entanglement-generation at a rate that depends only linearly on the transmissivity $\eta_{\rm}$ of a single link in the network, whereas the rate achieved by a linear repeater chain connecting Alice and Bob along the shortest path would decay as $\eta_{\rm}^{n_{\rm SP}}$ where $n_{\rm SP}$ is the length of the shortest path. Even a linear repeater chain can attain a rate that is proportional to $\eta_{\rm}$, but that requires repeater nodes equipped with error-corrected quantum processors~\footnote{Achievability of $\eta$ ebits/mode on a linear repeater chain---with $\eta$ being the transmissivity of the minimum-transmissivity link---with unrestricted processing at the repeater nodes follows from the single-photon dual-rail encoded realization of the Ekert-91 protocol~\cite{1991.PRL.Ekert.QCrypt} together with ideal BSMs, thereby establishing that the rate achieved by a linear chain of ideal repeaters is superior to that attained without any repeater assistance. Azuma {\em et al.} generalized the TGW bound to an upper bound on the rate achievable for a single entanglement generation flow (line repeater being a special case), which established $\log_2[(1+\eta)/(1-\eta)] \approx 2.89\eta$ for $\eta \ll 1$ as an upper bound to the rate~\cite{2016.NatComm.Azuma-Lo.QNetrateloss}. Pirandola generalized the PLOB upper bound~\cite{2016.ArXiv.Pirandola.CapRepQC} to repeater networks and proved a matching lower bound, showing that with an information-theoretic description of quantum repeaters, i.e., that are ideal fully-error-corrected universal quantum processors, that the optimal rate attainable for entanglement generation on a linear repeater chain is given by $\log_2[1/(1-\eta)] \approx 1.44\eta$ for $\eta \ll 1$.}. We achieve the same feature (rate proportionality to $\eta$) by multipath routing with percolation, with a much simpler repeater. We also analyze repeater protocols to support multiple entanglement-generation flows. This analysis reveals that simple space-time-division multiplexing strategies that use local link-state knowledge at nodes can outperform the best rate-region (the set of rates simultaneously achievable by different flows) attainable by repeater nodes that simply time share among assisting individual flows.

Our work also opens up a number of new questions that remain unanswered. We abstract off the entanglement routing problem to the following parameters: $G$ (network topology), $p$ (probability of successful creation of an entangled pair across one link in a given time step), $q$ (probability of a successful Bell measurement when attempted), $S$ (number of parallel links across a network edge) and $T$ (the number of time slots that a memory can hold a qubit before it decoheres).
Even in this simplified model, finding the rate-region optimizing routing protocol remains open. The aforesaid abstraction applies when the only source of imperfection at each component (including the quantum memories) is pure loss. Since our protocol only requires a quantum memory to hold a qubit for one entanglement attempt between neighboring stations ($T=1$), photon loss would indeed be the major source of imperfection in many implementations of the protocol. Accounting for more general errors would require purification of entanglement~\cite{1996.PRL.Deutsch-Sanpera.PrivAmp, 1996.PRL.Bennett-Wootters.Purification, 1998.PRL.Briegel-Zoller.BDCZ}, which will require us to introduce the Fidelity of entanglement during intermediate steps of the routing protocol as an additional parameter, as was done by Jiang {\em et al.}~\cite{2009.PRA.Jiang-Lukin.3genQR}. Furthermore, we restricted our analysis only to 2-qubit measurements at repeater nodes. Multi-qubit unitary operations and multi-qubit measurements at repeater nodes (e.g., a 3-qubit GHZ projection across three locally held qubits) would require more complex repeaters than those in repeater chains, but may improve the achievable rates. Finally, it will be interesting to consider repeater protocols for distillation of multi-partite entanglement shared between more than two parties, and a repeater network that supports multiple simultaneous flows of multi-partite entanglement generation.

\section{Background}

Let us consider a quantum network with topology described by a graph $G(V, E)$. Each of the $N = |V|$ nodes is equipped with a quantum repeater, and each of the $M = |E|$ edges is a lossy optical channel of range $L_i$ (km) and power transmissivity $\eta_i \propto e^{-\alpha L_i}$, $i \in E$. Consider $K$ source-destination (Alice-Bob) pairs $(A_j, B_j)$, $1 \le j \le K$, situated at (not necessarily distinct) nodes in $V$, each of which would like to generate maximally-entangled qubits (i.e., ebits) between themselves (and thus by definition not entangled with any other party, due to the monogamy property of entanglement), at the maximum rates possible $R_j$ (ebits per channel use). The high-level objective is: {\em Given a class of quantum and classical operations at each of the repeater nodes of the underlying network, what operations should be performed at the repeater stations to maximize the rate region $(R_1, R_2, \ldots, R_K)$ simultaneously achievable by the entanglement flows?} More importantly, one would like to address networking questions such as: (a) what is the maximum rate-region attainable, (b) what is the tradeoff between sum throughput and latency of the $K$ entanglement flows,  and (c) where should repeater nodes be placed, with constraints on devices (e.g., memories, sources, and detectors), to maximize the attainable rate region; all being subject to various practical considerations. Ultimately one would like to develop explicit and efficient practical quantum routing protocols that employ quantum operations implemented via lossy and noisy devices, while only requiring local link-state knowledge and limited knowledge of the global network topology, analogous to the classical internet.

The entanglement-generation rate across a link of transmissivity $\eta$, in the absence of any repeater mediation, is limited to $-\log_2(1-\eta)$ ebits per mode, amounting to $\approx 1.44\eta$ ebits per mode when $\eta \ll 1$~\cite{2017.NatComm.Pirandola-Banchi.QuantCommUltRate}\footnote{The achievability of $-\log_2(1-\eta)$ ebits per mode of secret communication rate over the lossy channel (with two way authenticated public classical communication) was first proven in 2009 by Pirandola {\em et al.}~\cite{2009.PRL.Pirandola-Lloyd.SKCLB}. In 2014, Takeoka {\em et al.} proved an upper bound to the secret-key agreement capacity, $\log_2[(1+\eta)/(1-\eta)]$ ebits per mode~\cite{2014.NatureComm.Takeoka-Wilde.TGWBound}, which equals $\approx 2.88\eta$ ebits per mode when $\eta \ll 1$, thereby establishing that the rate attained by {\em any} protocol must decay linearly with the channel's transmissivity and hence exponentially with distance $L$ in optical fiber (since $\eta \sim e^{\alpha L}$). In 2015, Pirandola {\em et al.} proved an improved (weak converse) upper bound of $-\log_2(1-\eta)$ ebits per mode, which established that as the secret key agreement capacity of the pure loss bosonic channel~\cite{2017.NatComm.Pirandola-Banchi.QuantCommUltRate}. In 2017, Wilde {\em et al.} proved $-\log_2(1-\eta)$ ebits per mode as a strong converse upper bound to the secret-key agreement capacity~\cite{2017.IEEETransInfTheo.Wilde-Berta.PricComQChanStrongConv}.}. The number of modes per second is a device-dependent constant, upper bounded by the maximum of the optical bandwidth of the source and the electrical bandwidth of the detector. Since $\eta  \sim e^{-\alpha L}$ where $L$ is the length of optical fiber, the ebits-per-mode rate also falls off exponentially with range $L$. Most analyses of repeater networks have been limited to linear chains, with the objective of outperforming the repeater-less bound~\cite{1998.PRL.Briegel-Zoller.BDCZ, 2009.PRA.Jiang-Lukin.3genQR, 2016.SciRep.Muralidharan-Jiang.RepeaterGen, 2014.PRL.Sinclair-Tittel.AtomFreqCombRep, 2015.PRA.Guha-Tittel.QRRateLossAnalysis, 2015.NatureComm.Azuma-Lo.AllOptRep, 2017.PRA.Pant-Guha.AllOptRepRescources, 2016.PRL.Ewert-VanLoock.QPCPhotRep}. Pirandola analyzed entanglement-generation capacities of repeater networks assuming ideal repeater nodes, i.e., those equipped with fully-error-corrected quantum processors and argued that for a single flow ($K = 1$), the maximum entanglement-generation rate $R_1$ reduces to the classical max-flow min-cut problem with edge $e$ being associated with capacity $C(e) = -\log_2(1 - \eta(e))$ ebits per channel use~\cite{2016.ArXiv.Pirandola.CapRepQC}, where $\eta(e)$ is the transmissivity of edge $e$. Pirandola subsequently argued that classical cut-set bounds with the above link capacity give outer bounds to the $K$-flow capacity region, but again, for ideal repeater nodes. Azuma {\em et al.} independently established an upper bound~\cite{2016.NatComm.Azuma-Lo.QNetrateloss} to the entanglement generation bound to the rate which has the same asymptotic scaling but is not tight. Azuma has also looked at an ``aggregated" protocol in which repeater protocols run in parallel~\cite{2016.ArXiv.Azuma.AggQuantRep}. Schoute and co-authors~\cite{2016.ArXiv.Schoute-Wehner.PerfQnetroute} developed routing protocols on specific network topologies and found scaling laws as functions of $N$, the number of qubits in the memories at nodes, and the time and space consumed by the routing algorithms, under the assumption that each link generates a perfect, lossless EPR pair in every time slot, and that the nodes' actions are limited to (perfect) Bell-state measurements (BSMs). Ac\'{i}n and co-authors~\cite{2007.NaturePhys.Acin-Lewenstein.EntgPerc} have considered the problem of entanglement percolation where neighboring nodes share a perfect, lossless pure state. Further, van Meter and co-authors developed explicit networking protocols also restricted to pair-wise EPR pair generation and BSMs, but accounting for imperfect fidelities of the EPR pairs (and thus requiring purification over multiple imperfect pairs), and finite coherence times of the qubit memories~\cite{2014.Book.VanMeter.QuantumNetworking}. There has also been previous work on quantum network coding~\cite{2007.STACS.Hayashi-Yamashita.QNetCoding, 2009.Springer.Kobayashi-Roetteler.QnetCodewithClassComm, 2012.PRA.Satoh-Imai.QuantNetCodeQRep, 2016.PRA.Satoh-VanMeter.QNetCodeAn} and linear-optic quantum routers~\cite{2013.PRA.Lemretal.LinOptRout}.

\begin{figure}[t]
\includegraphics[width=0.5\textwidth]{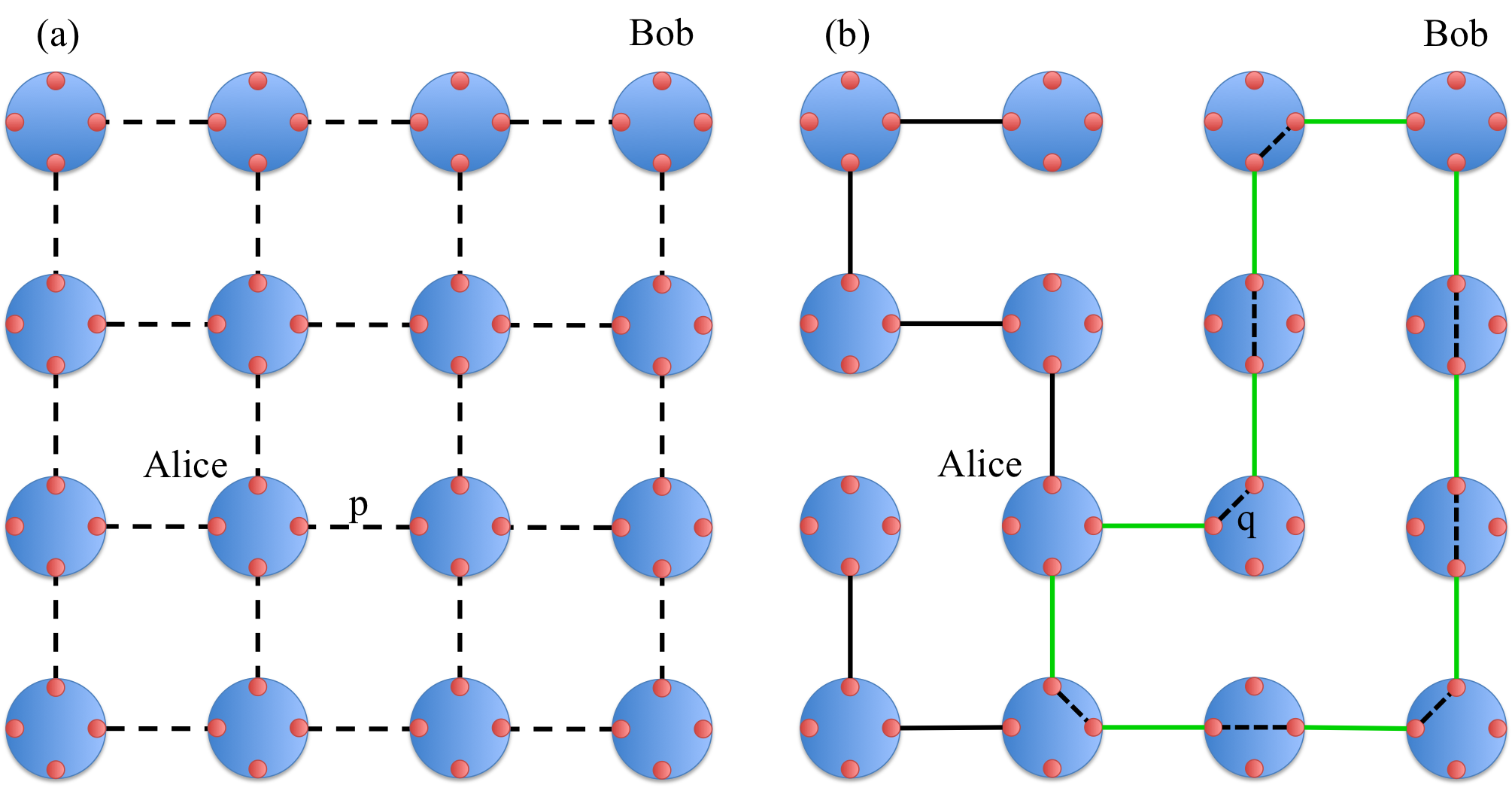}
\caption{Schematic of a square-grid topology. The blue circles represent repeater stations and the red circles represent quantum memories. Every cycle (time slot) of the protocol consists of two phases. (a) In the first (external) phase, entanglement is attempted between neighboring repeaters along all edges, each of which succeed with probability $p$ (dashed lines). (b) In the second (internal) phase, entanglement swaps are attempted within each repeater node based on the successes and failures of the neighboring links in the first phase---with the objective of creating an unbroken end-to-end connection between Alice and Bob. Each of these internal connections succeed with probability $q$. Memories can hold qubits for $T \ge 1$ time slots.}
\centering
\label{fig: Qnetworkschematic}
\end{figure}

\section{Entanglement routing protocols}

\subsection{Problem statement and notation}

%We consider the general repeater network setting schematically depicted in Fig.~\ref{fig:USnetwork}, drawn for illustrative purposes at a continental scale. While our protocol can be adapted to any network topology, our results in this paper will be evaluated for a 2D square-grid topology, shown in Fig.~\ref{fig: Qnetworkschematic}. 

Consider a graph $G(V, E)$ that denotes the topology of the repeater network. See Fig.~\ref{fig:USnetwork} for an illustration. Each node $v \in V$ is a repeater (blue circles), and each edge $e \in E$ is a physical link connecting two repeater nodes. $S(e) \in {\mathbb Z}^+$ is an integer edge weight, which corresponds to the number of parallel (single spatial, spectral or polarization mode) channels across the edge $e$ (shown using blue lines). The number of memories at node $v$ is  $\sum_{e \in {\cal N}(v)} S(e)$ (see Fig.~\ref{fig:USnetwork}), where the sum is over ${\cal N}(v)$, the set of nearest neighbor edges of $v$, with $d(v) = |{\cal N}(v)|$ the degree of node $v$.

Time is slotted. We assume that each memory can hold a qubit perfectly for $T \ge 1$ time slots (T should be much smaller than the memory's coherence time). Each time slot $t$, $t = 1, 2, \ldots$, is divided into two phases: the ``external" phase and the ``internal" phase, which occur in that order. During the external phase, each of the $S(e)$ pairs of memories across an edge $e$ attempts to establish a shared entangled (EPR) pair. An entanglement attempt across any one of the $S(e)$ parallel links across edge $e$ succeeds with probability $p_0(e) \sim \eta(e)$~\cite{2014.NatureComm.Takeoka-Wilde.TGWBound, 2017.NatComm.Pirandola-Banchi.QuantCommUltRate}, where $\eta(e) \sim e^{-\alpha L(e)}$ is the transmissivity of a lossy optical channel of length $L(e)$. Using two-way classical communication over edge $e(u,v)$, neighboring repeater nodes $u, v$ learn which of the $S(e)$ parallel links (if any) succeeded in the external phase. 

Let us assume that neighboring repeaters pick upto one successfully-created ebit (i.e., ignore multiple successes if any) as in Ref.~\cite{2015.PRA.Guha-Tittel.QRRateLossAnalysis, 2017.PRA.Pant-Guha.AllOptRepRescources}, in which case the probability that one ebit is established successfully across the edge $e$ during the external phase is given by: $p(e) = 1 - (1-p_0)^{S(e)}$. Let us also assume $S(e) = S, \forall e \in E$, which in turn gives us $p(e) = p, \forall e \in E$. While our results in this paper can be adapted to any network topology, we will henceforth use the 2D regular square grid topology (Fig.~\ref{fig: Qnetworkschematic}) to illustrate the performance of our routing algorithms.

One instance of the resulting external links created between repeater nodes after the external phase is shown in Fig.~\ref{fig: Qnetworkschematic}(b) using solid lines. In the internal phase, entanglement swap (BSM) operations are attempted locally at each repeater node between pairs of qubit memories. We associate these BSM attempts as {\em internal links}, i.e. links between memories internal to a repeater node, shown using dotted lines inside repeaters in Fig.~\ref{fig: Qnetworkschematic}(b). If $T > 1$, a repeater node can attempt a BSM between qubits held in two memories that were entangled with their respective neighboring node's qubits in two different time slots. For minimizing the demands on memory coherence time~\cite{2015.PRA.Guha-Tittel.QRRateLossAnalysis, 2017.PRA.Pant-Guha.AllOptRepRescources}, we will assume $T=1$. So, BSMs will always be attempted between two qubits in distinct memories that were entangled with their respective counterparts at their respective neighboring nodes in the same time slot. Each of these internal-link attempts succeed with probability $q$. Therefore, after the conclusion of one time slot, along a path comprising $k$ edges (and thus $k-1$ repeater nodes), one ebit is successfully shared between the end points of the path with probability $p^kq^{k-1}$. The maximum number of ebits that can be shared between Alice (say, node $a$) and Bob (say, node $b$) after one time slot is ${\rm min}\left\{d(a), d(b)\right\}$, assuming $S$ is the same over all edges. For the square-grid topology shown, the maximum number of ebits that can be generated between Alice and Bob in each time slot is $4$. 

The remainder of the paper is dedicated to finding the optimal strategy for each repeater node in order to decide which locally-held qubits to attempt BSM(s) on during the internal phase of a time slot, based ideally only on knowledge of the outcomes (success or failure) of the nearest neighbor links, i.e., local link-state knowledge, during the respective preceding external phases. We will assume that each repeater node is aware of the overall network topology as well as the locations of the $K$ Alice-Bob pairs. The goal of the optimal repeater strategy will be to attain the maximum entanglement-generation rate (if there is a single Alice and Bob, i.e., $K=1$) or the maximum rate-region for multiple flows (i.e., $K>1$).

%In the physical implementation of the protocol, $p$ could be increased by multiplexing entanglement generation across $n$ attempts between $n$ different memory elements and picking one successfully-created ebit, as is common in some linear repeater chains that have been analyzed~\cite{2015.PRA.Guha-Tittel.QRRateLossAnalysis, 2017.PRA.Pant-Guha.AllOptRepRescources}. If the probability of entangling two memories in neighboring repeater links is $p_0$, the probability creating one or more external edge would be $p = 1 - (1-p_0)^n$. We will henceforth use $p$ for the edge probability $p$ without going into the details of how it might have been enhanced by multiplexing, and more importantly ignoring the prospect of using multiple successfully created ebits across an edge in one clock cycle to route entanglement across the network. Using multiple successfully created ebits across an edge may not lead to a huge improvement unless the probability of creating atleast one link is close to one, because in this regime, the probability of sharing atleast two ebits across an edge $p^{(2)}$ is significantly smaller than $p$. Furthermore, using multiple ebits between neighboring repeater stations does not improve the performance of a linear repeater chain if the repeater spacing is optimized [CITE Saikat's unpublished paper].

\subsection{Multipath routing of a single entanglement flow}

\subsubsection{Entanglement routing with global link-state information}

We begin with the assumption that global link-state knowledge is available at each repeater node, i.e., the state of every external link in the network after the external phase is known to every repeater in the network and can be used to determine the choice of which internal links to attempt within the nodes. Each memory can only be part of one entanglement swap, i.e., each red node can only be part of one internal edge. Consider the following greedy algorithm to choose the internal links: consider the subgraph induced by the successful external links and the repeater nodes (at the end of the external phase), and find in it the shortest path connecting Alice and Bob. If no connected path between Alice and Bob exists, no shared ebits are generated in that time slot. If a shortest path of length $k_1$ is found, all internal links along the nodes of that path are attempted, and the probability a shared ebit is generated by this path is the probability that all $k_1-1$ internal link attempts were successful, i.e., $q^{k_1-1}$. We then remove all the (external and internal) links of the above path from the subgraph, and find a shortest path connecting Alice and Bob in the pruned subgraph. Note that instead of removing the links of the first path from the subgraph, we could simply search for a shortest path in the original subgraph but one that is edge disjoint from the previous path. If such a path exists, we again attempt all internal links at the nodes of this path, so the probability the path contributes to the generation of an ebit between Alice and Bob is $q^{k_2-1}$ where $k_2$ is the length of the second path; and so on. 

The entanglement generation rate achieved using this greedy algorithm $R_g$ is the sum of expected rates (in ebits per time slot) from these paths. Given the degree-$4$ nodes in a square grid topology, there can be a maximum of four edge disjoint paths between Alice and Bob. Fig.~\ref{fig: Qnetworkschematic}(b) illustrates our greedy algorithm. Given the set of external links created, the shortest path has length $k_1 = 4$, the next path has length $k_2 = 6$, and no further paths can be found. The two edge-disjoint paths are highlighted in green. Hence, the internal links depicted with the dashed lines in Fig.~\ref{fig: Qnetworkschematic}(b) are attempted and the expected number of shared ebits generated in this time cycle is: $q^{k_1 - 1} + q^{k_2 - 1}$. The net entanglement generation rate is the expectation of sums like the above (with up to four terms) over many random instantiations of the $(p, 1-p)$ external-link creations during the external phase of many time slots. Evaluating this expected rate $R_{\rm g}(p,q)$ achieved by the above routing strategy analytically as a function of the Alice-Bob distance ($X_1, X_2$) is difficult, even for a square-grid topology.

The intuition behind this simple greedy algorithm is that the entanglement generation rate along a path of length $k$ decays exponentially as $q^{k-1}$, suggesting that attempting internal links to facilitate connections along the shortest path first would optimize the expected rate. However, it is possible to draw random instances of successes of external links, where either one of the two possible options---(1) picking the shortest path (which disrupts all other paths) and (2) picking two edge disjoint (but longer) paths---could yield either a larger or a smaller expected rate than the other, depending upon the value of $q$. If $q$ is larger than a threshold, option (2) would have a larger expected rate and vice versa. Finding the global optimal rule remains an open problem. It is easy however to prove that the greedy algorithm achieves a rate within a factor of four of the optimum algorithm employing global link-state knowledge, $R_{\rm{opt}}(p,q)$. Let us denote the length of the shortest path between Alice and Bob with Manhattan distance $(X, Y)$ in the induced subgraph after the external phase, as $n_{\rm SP}(p)$. This quantity is of interest in percolation theory, and is not completely understood analytically. It undergoes a sharp transition (i.e., starts out large and suddenly jumps to a much smaller value) as $p$ crosses $p_c$ from below to above. Clearly, $R_{\rm g}(p,q) \geq {\mathbb{E}}[q^{n_{\rm SP}(p) - 1}]$ since using the shortest path is the first step of the greedy algorithm. Furthermore, since the optimal rule can create entanglement over a maximum of four edge-disjoint paths in each time step, each of which must have a length no less than the length of the shortest path, $R_{\rm opt}(p,q) \leq {\mathbb{E}}[4q^{n_{\rm SP}(p) - 1}] \triangleq R^{\rm{(UB)}}_{\rm opt}(p,q)$. Therefore, $R_{\rm opt} \geq R_{\rm g} \geq R_{\rm opt}/4$, i.e., the greedy rule will achieve the same rate-vs.-distance scaling as the optimal algorithm that employs global link-state knowledge, and can be worse only by a small constant factor.

\begin{figure*}[t]
\includegraphics[width=0.9\textwidth]{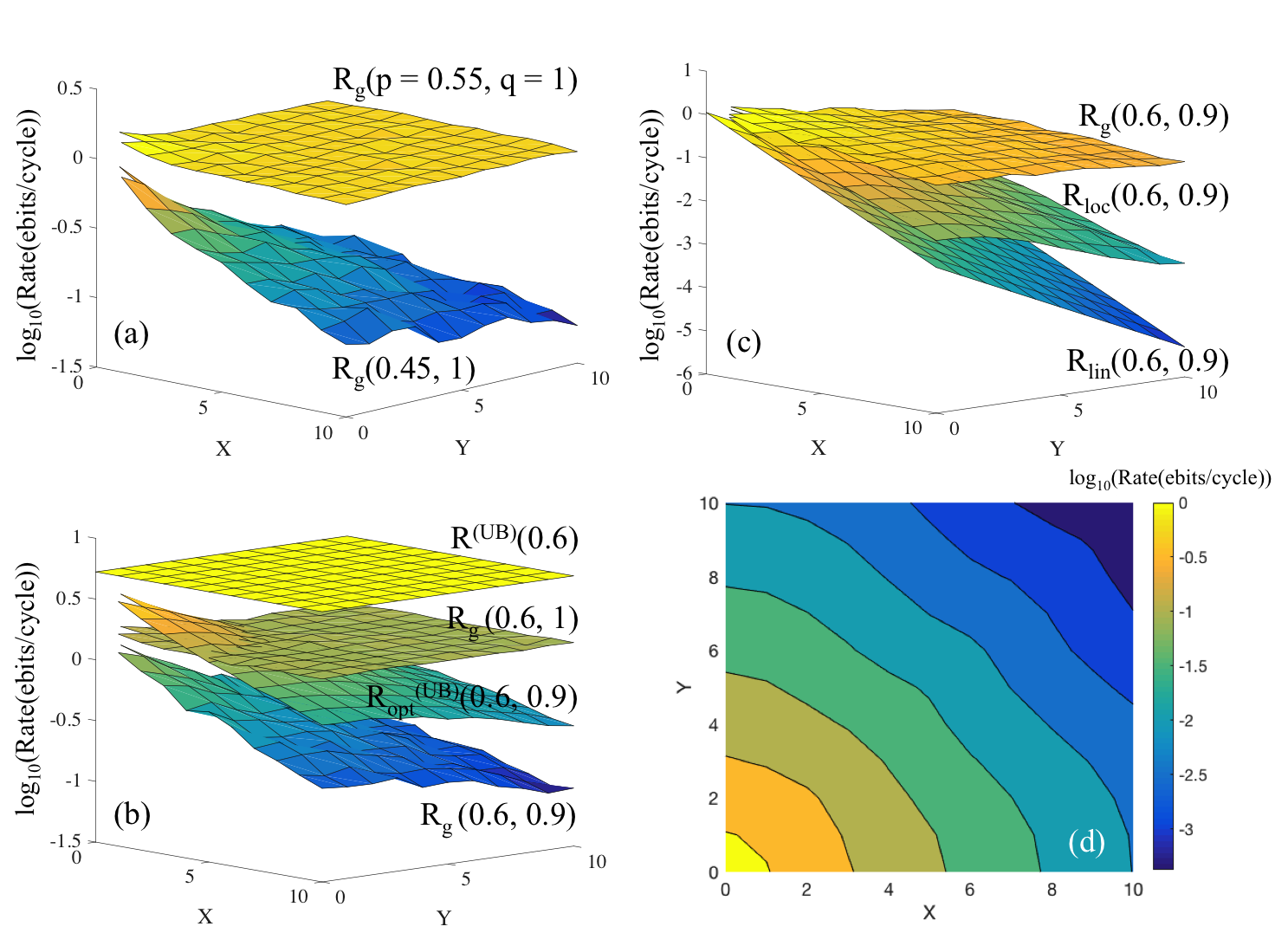}
\caption{Entanglement generation rate as a function of the Alice-Bob separation along $X$ and $Y$ (on a square grid) as a function of ($p, q$); (a) $R_{\rm g}(p,q)$ is the rate attained by a global-knowledge-based protocol we propose where each node, in each time step, knows whether any link in the entire network succeeded or failed to establish entanglement. For the case of $q = 1$, $R_{\rm g}$ is distance independent when $p$ is greater than the bond percolation threshold (0.5 for the square lattice) and decays exponentially if it is below the threshold. (b) $R^{\rm{(UB)}}$(0.6) is the distance-independent Pirandola rate upper bound for $p = 0.6$, achieving which requires perfect quantum processing at repeater nodes. $R_{\rm g}(0.6, 1)$ is also distance independent, and within a factor ~3.6 of $R^{\rm{(UB)}}(0.6)$. With $q < 1$, e.g., $R_g(0.6, 0.9)$, the rate decays exponentially with distance. $R_{\rm opt}^{\rm{(UB)}}$ is an upper bound on the rate attainable with global-knowledge by any protocol. (c) $R_{\rm loc}$ is attained by a protocol we propose where each node, in each time step, only needs to know the link state of neighboring edges. The rate-distance scaling exponent of $R_{\rm loc}$ is clearly worse than $R_{\rm g}$, but is significantly superior to that of a linear repeater chain along the shortest path, $R_{\rm lin}$, demonstrating multi-path routing advantage even with local link-state knowledge. (d) Contour plot of the entanglement generation rate with the local rule when $p = 0.6$ and $q = 0.9$. Although the Alice to Bob distance along the network links is $X + Y$, there is a noticeable enhancement in the rate along the $X=Y$ direction because of more Alice-Bob paths of similar length.}
\centering
\label{fig: Rateplots}
\end{figure*}

In Fig.~\ref{fig: Rateplots}(a) we plot $R_{\rm g}(p, q)$ as a function of the Alice-Bob Manhattan distance ($X, Y$) on the square grid (measured in number of hops) with $q = 1$. When $p > p_c$, the bond percolation threshold of the underlying network ($p_c = 0.5$ for the square lattice), a {\em giant connected component} is formed by the external links alone at the end of the first (external) phase of a time slot. Recall that the rate along a length $k$ path is $p^kq^{k-1}$, where $p \sim \eta$ is the transmissivity of each link. In the network case, when $p>p_c$ and $q = 1$, the $p^k$ portion of the rate expression becomes immaterial for scaling with Alice-Bob distance, since percolation guarantees a connected path to exist between Alice and Bob along successful external links in each time slot. So, if $q=1$, $R_{\rm g}(p,q)$ remains essentially distance invariant. When $p < p_c$, the rate falls off exponentially with distance (even when $q=1$). It is instructive to note here that the optimal rate (entanglement-generation capacity) achievable on a single length $k$ path does not depend on $k$, and only on the transmissivity of the lossiest link in the path, i.e., $C \sim \eta$~\cite{2016.ArXiv.Pirandola.CapRepQC}, but achieving this requires infinite-coherence-time quantum memories and ideal quantum operations at nodes. The multi-path gain in the $p > p_c$ regime lets us achieve a distance-independent rate, but with memories whose coherence times are no more than one time slot. The rates are calculated using monte-carlo simulations which results in some numerical noise that is insignificant compared to the difference between the plots, but is visible in $R_{\rm g}(0.45,1)$.

A general upper bound on the entanglement generation rate is given by the min-cut of the graph~\cite{2016.ArXiv.Pirandola.CapRepQC}, and for a square lattice, is given by $R^{\rm{(UB)}}(p) = -\log_2[(1-p)^4]$. $R^{\rm{(UB)}}(0.6)$ is plotted in Fig.~\ref{fig: Rateplots}(b). The known methods for achieving $R^{\rm{(UB)}}$ require infinite coherence time memories and error-corrected quantum processors at each node. For our implementation (assuming global link state knowledge), $R_{\rm g}(0.6, 1)$ is also plotted in Fig.~\ref{fig: Rateplots}(b). Although our protocol only requires memories to hold entanglement for one time step, the multi path advantage gives us the same constant rate-distance scaling and within a factor of $\sim 3.6$ of $R^{\rm{(UB)}}(0.6)$. The assumption of perfect BSMs is unrealistic and $q < 1$, in which case $R_g(p, q)$ falls off exponentially with distance; even when $p > p_c$, as seen in the plot for $R_{\rm g}(0.6, 0.9)$. Finally, we plot the above discussed upper bound on $R_{\rm g}$, $R_{\rm opt}^{\rm{(UB)}}(0.6,0.9)$, which as expected has the rate-distance scaling of $R_{\rm g}$, but larger by a factor less than $4$.

%Clearly, we can do better by attempting shared entanglement along multiple paths.  Since there are four memories at Alice and Bob, there can be a maximum of four possible paths from Alice to Bob in each cycle and the probability of success of each path is less than or equal to $q^{k_{m}-1}$. Hence, we can establish an upper bound to the optimum communication protocol: $R_{UB} = 4q^{k_{m}-1}$ ebits/cycle $= 4R_{LB}$. Since the lower and upper bounds only differ by a factor of 4, these bounds are sufficient for capturing the scaling of the protocol. 

\subsubsection{Entanglement routing with local link-state information}
\begin{figure}[t]
\includegraphics[width=0.5\textwidth]{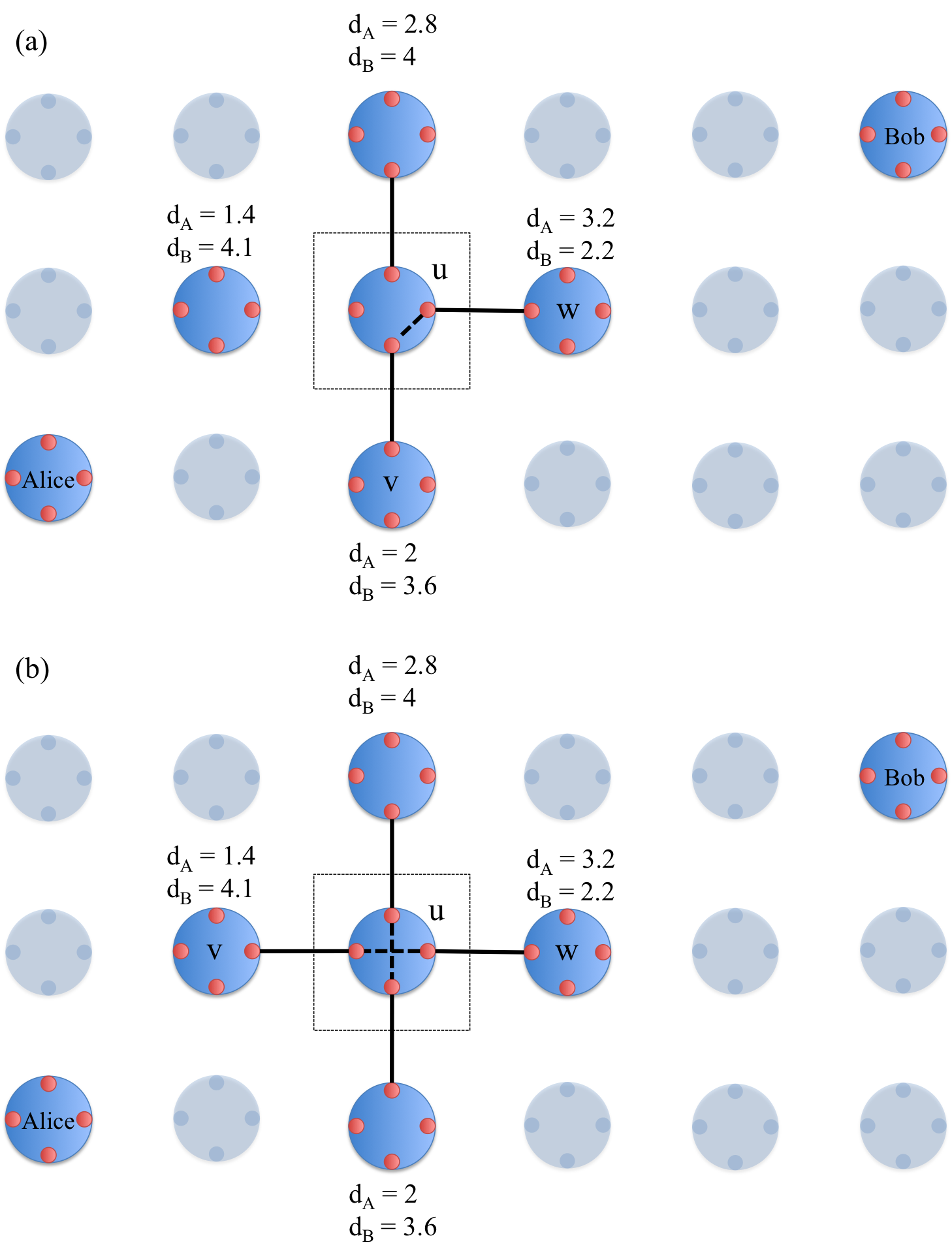}
\caption{The entanglement swap rule used at the repeater C in the dotted box in the case of local link-state knowledge. A and B are the repeaters closest to Alice and Bob, respectively, with a direct edge to C. (a) If two or three links are up, the memories linked to A and B undergo an entanglement swap. (b) If four links are up, the remaining two memories also undergo an entanglement swap.}
\centering
\label{fig: Qnetwork_localrule}
\end{figure}

$R_{\rm g}(p, q)$, the rate attained by the protocol described in the previous subsection that employs global link-state knowledge, is re-plotted in Fig.~\ref{fig: Rateplots}(c). We also plot $R_{\rm lin}(p,q) = p^{n_{\rm SP}(1)}q^{{n_{\rm SP}(1)}-1}$, the rate attained by a single linear repeater chain, where ${n_{\rm SP}(1)}$ is the shortest-path length between Alice and Bob along the edges of the underlying square grid. The assumption of global link-state knowledge in large networks is unrealistic, as it requires memories whose coherence time increases with the network size due to the time required for the traversal of link-state information across the entire network. In this section, we describe a more realistic protocol in which knowledge of success and failure of an external link at each time slot is communicated only to the two repeater nodes connected by the link, as is the case in the analysis of many `second-generation' linear repeater chains~\cite{2016.SciRep.Muralidharan-Jiang.RepeaterGen, 2015.PRA.Guha-Tittel.QRRateLossAnalysis,2009.PRA.Jiang-Lukin.3genQR}. Repeater nodes need to decide on which pair(s) of memories BSMs should be attempted (i.e., which internal links to attempt), based only on information about the states of external links adjacent to them. We assume that network topology and positions of Alice and Bob are known to each repeater station, and communicated classically beforehand. 

Let us consider a local repeater rule illustrated in Fig.~\ref{fig: Qnetwork_localrule}. The repeater $u$ inside the dotted square box has to make a decision regarding which internal edges to attempt based on the information of which of the four neighboring external edges have been successfully created in the external phase. We associate $d_A$ and $d_B$ as the distance to Alice and Bob, respectively, at every repeater node. We use the ${\mathbb{L}}^2$ norm distance to Alice (resp. Bob) as $d_A$ (resp. $d_B$). Of all the nearest neighbor nodes of $u$ whose links to $u$ were successful in that time slot, we label the one that has the minimum $d_A$ as $v$. Similarly, the neighbor with a successful external link with $u$ and the minimum $d_B$ is labelled $w$. An internal link is attempted between the memories connected to $v$ and $w$ respectively, as shown in Fig.~\ref{fig: Qnetwork_localrule}(a). If $v$ and $w$ are the same node, $v$ (or $w$) is replaced by node $u$'s nearest-neighbor node with the next smallest value of $d_A$ (or $d_B$). The choice of whether to replace $v$ or $w$ is made in a manner that minimizes the sum of $d_A$ and $d_B$ from the eventually chosen two neighbors to connect. If all four external links are successful, an additional internal link is attempted between the remaining two memories as shown in Fig.~\ref{fig: Qnetwork_localrule}(b). If only one of the neighboring external links is successful, no internal links are attempted, since this repeater node cannot be part of a path from Alice to Bob in that time slot. If two neighbors have the same values of $d_A$ and $d_B$, an unbiased coin is tossed to determine the choice of $v$ and $w$, to preserve symmetry in the protocol.

The entanglement generation rate $R_{\rm loc}(p,q)$ achieved by the above described local rule is plotted in Fig.~\ref{fig: Rateplots}(c) and compared to $R_{\rm g}(p,q)$ and $R_{\rm lin}(p,q)$. We use $p = 0.6$ and $q = 0.9$, the same values used for the global-information rate plots in Fig.~\ref{fig: Rateplots}(b). As one expects, the rate-distance scaling of $R_{\rm loc}$ is worse than that of $R_{\rm g}$. However, the rate-distance scaling exponent achieved by the local rule is superior to that of a linear chain, even though the physical elements employed to build the repeaters are identical. This is proven analytically in appendix~\ref{sec:localrateLB}. Note that each of the three rates $R_{\rm g}$, $R_{\rm loc}$, $R_{\rm lin}$ fall exponentially with distance, but the exponents are different. The scaling advantage of $R_{\rm loc}$ over $R_{\rm lin}$ arises because the local rule allows the entanglement-generation flow between Alice and Bob to find different (and potentially simultaneously multiple) paths in different time slots, and does not have to rely on all links along a linear chain to be successful. This is analogous to multi-path routing in a classical computer network. The contour plot in Fig.~\ref{fig: Rateplots}(d) further illustrates this point: there is a noticeable enhancement of $R_{\rm loc}$ along the $X=Y$ line because the diagonal direction contains the largest spatial density of possible paths between Alice and Bob. The scaling advantage over $R_{\rm lin}$ persists in any direction, i.e., along $Y=0$ as well. 

Sweeping over different values of $p$ and $q$, we find that the multi-path advantage relative to a linear repeater chain increases as $p$ decreases from unity, but there is little relative improvement as $q$ is varied (see Appendix~\ref{sec:multipath_app}). 

Clearly, other distance metrics (e.g., ${\mathbb{L}}^p$ norm for $p \ge 1$) can be used in lieu of the ${\mathbb{L}}^2$ norm in the algorithm described above. In Appendix~\ref{sec:recursive}, we present a recursive numerical evaluation technique to find the rate-optimal distance metric, which can be applied to any network topology. For planar network topologies, the ${\mathbb{L}}^2$ norm appears near-optimal for our local routing algorithm.

An analytical enumeration of the expected number of edge-disjoint paths as a function of $p$ between Alice and Bob separated by a given distance ($X, Y$) in a bond-percolation instance (i.e., with $p > p_c$) of a network is an open question, the solution of which will enable a firmer quantitative understanding of the multi-path advantage in entanglement generation in a repeater network.

%Sweeping over different values of $p$ and $q$, we find that the scaling of the rate $R_{loc}$ with the $n$, the shortest path between Alice and Bob, goes as $~ f^n(p,q)$. Since the rate of a linear repeater chain goes as $p^nq^{n-1}$, the relative improvement in scaling can be quantified by $f(p,q)/pq$ which is show  

\subsection{Simultaneous entanglement flows}

\begin{figure}
\includegraphics[width=\columnwidth]{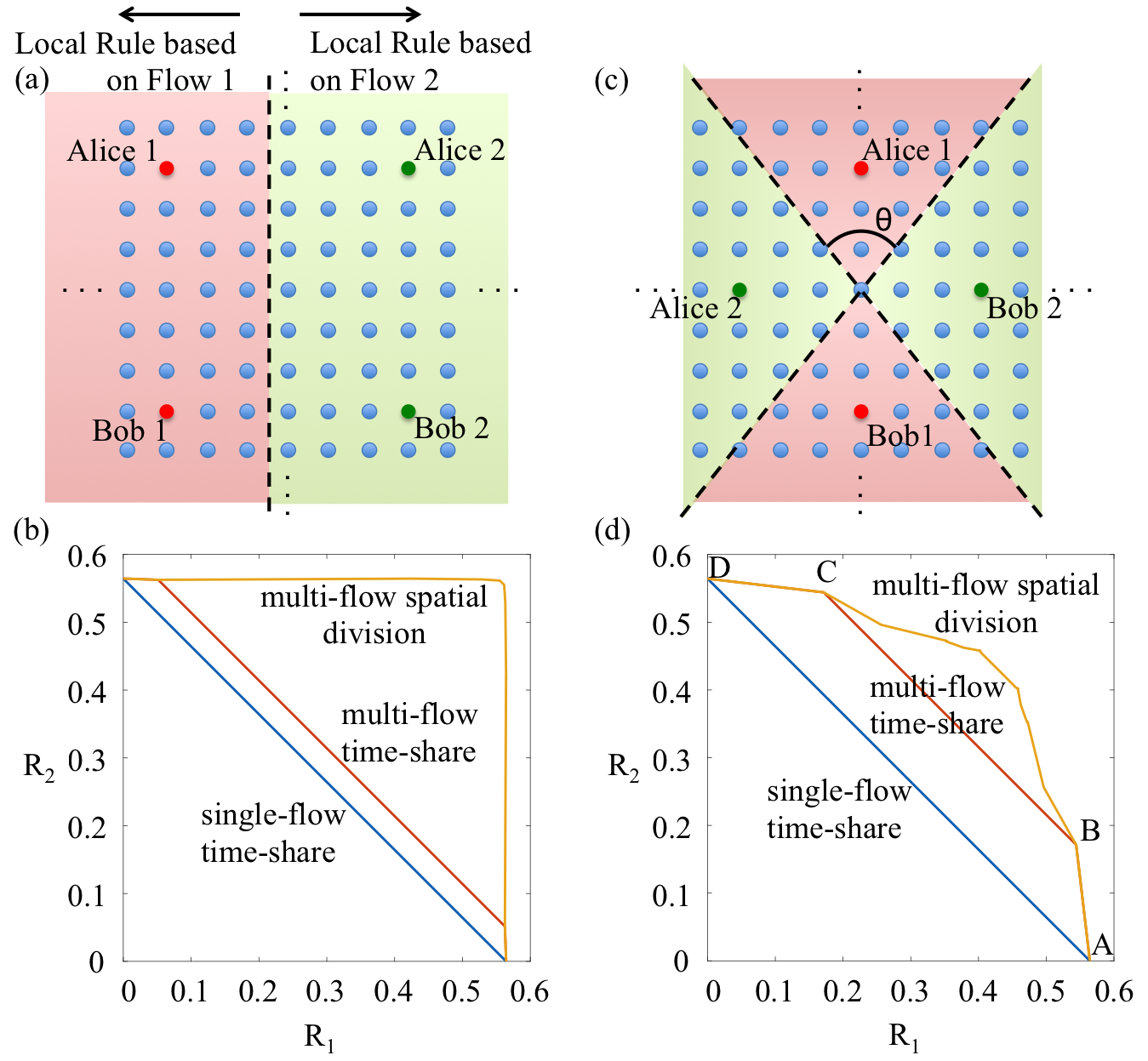}
\caption{(a) Multi-flow routing for two Alice-Bob pairs that lie along the sides of a $6 \times 6$ square, embedded in a $100 \times 100$ grid; (b) rate region ($R_1, R_2$) with different rules at repeater nodes, each employing local link-state knowledge, for $p = q = 0.9$. (c) Multi flow routing when the Alice-Bob paths cross (d) multiflow rate region for two local-knowledge rules.}
\centering
\label{fig: multiflow}
\end{figure}

In this section, we consider simultaneous entanglement-generation flows between two Alice-Bob pairs, using local link state knowledge at all repeater nodes. Consider two pairs Alice 1 - Bob 1 (red nodes) and Alice 2 - Bob 2 (green nodes) as shown in the two scenarios in Fig.~\ref{fig: multiflow}. In Fig.~\ref{fig: multiflow}(a), the shortest paths connecting the two Alice-Bob pairs do not cross, but they do in Fig.~\ref{fig: multiflow}(b). In both cases, they are placed at the four corners of a $6 \times 6$ square grid, embedded within a large square grid network. Denote by $R_1$ and $R_2$ the entanglement generation rates achieved by the two Alice-Bob pairs respectively. We first consider the case of non-intersecting flows shown in Fig.~\ref{fig: multiflow}(a). A simple strategy is for every single repeater node (including the nodes labeled as the two Alices and Bobs) to use the local rule described in the previous section tailored to support the Alice 1-Bob 1 flow for a fraction, $\lambda$, of the time slots and to support the Alice 2-Bob 2 flow for the remaining $1-\lambda$ fraction. For $p = q = 0.9$, the rate region attained by varying $\lambda \in [0,1]$ is depicted with the blue line in Fig.~\ref{fig: multiflow}(b), which we refer to as single-flow time-share. However, if every repeater with the exception of the Alices and Bobs carry out the above time-sharing strategy, even when all repeater nodes support flow 1, there is still some `left-over' non-zero $R_2$ that is attained. This multi-flow time-share rate region is shown using the red line in Fig.~\ref{fig: multiflow}(a).

\begin{figure}
\includegraphics[width=\columnwidth]{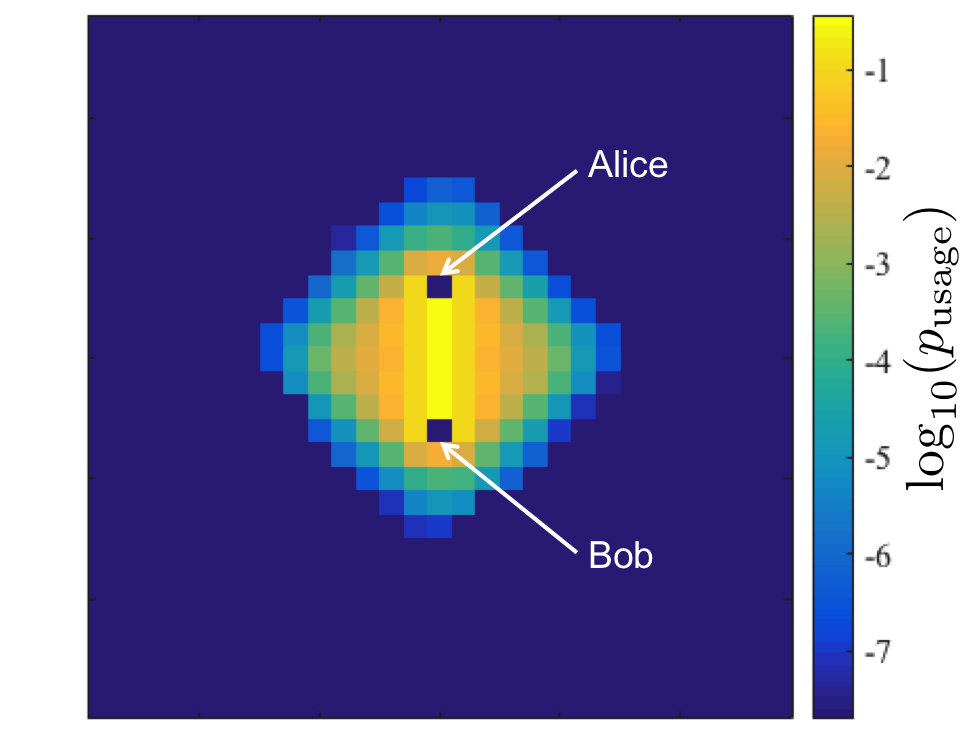}
\caption{A heat map plotting $p_{\rm usage}$, the probability that a given repeater node is involved in a successful creation of a shared ebit generated between Alice and Bob, separated by $6$ hops in an underlying square grid topology, when our local rule is employed. We assume $p=0.9$ and $q=0.9$.}
\centering
\label{fig:heatmap}
\end{figure}
In Fig.~\ref{fig:heatmap}, for the case that Alice and Bob are separated by $6$ hops on the square grid, we plot a color map of $p_{\rm usage}$, the probability a given repeater node is involved in a successful creation of a shared ebit generated between Alice and Bob when our local rule is employed. We observe that only the repeaters lying in a small spatial region surrounding the straight line joining Alice and Bob are used significantly. This observation motivates a multi-flow spatial-division rule, in which we divide the network between the two flows, as shown in Fig.~\ref{fig: multiflow}(a). Any repeater in the red shaded region follows the local rule tied to the Alice 1 - Bob 1 flow while repeaters in the green region operate with the local rule tied to the Alice 2 - Bob 2 flow. The placement of the boundary determines the rates $R_1$ and $R_2$. The rate region attained is plotted with the yellow line in Fig.~\ref{fig: multiflow}(b). This significantly outperforms time sharing. The two flows can co-exist and operate with a very small reduction from their individual best rates, because the repeaters they respectively benefit from the most form almost disjoint sets.

In the other extreme, we consider two Alice-Bob pairs, still separated by six hops, but with their shortest paths crossing as shown in Fig.~\ref{fig: multiflow}(c). The rate region attained by multi-flow time sharing, shown by the line segment BC, still provides an improvement over single-flow time-sharing, shown by the line segment AD, as shown in Fig.~\ref{fig: multiflow}(d). It is interesting to note that the maximum $R_1$ under multi-flow time sharing (point B) is slightly lower than maximum $R_1$ with the single-flow time-share rule (point A). This happens because unlike in single flow time-share, the nodes at Alice 2 and Bob 2 do not contribute to $R_1$ under multi-flow time-share. A point along AB represents time sharing between the strategies at points A and B. To further increase the rate, we adopt a multi-flow spatial division strategy in which nodes in the red region are configured to assist flow 1 and nodes in the green region are configured to assist flow 2. Varying the angle $\theta$ demarcating those regions results in the rate region shown by the yellow line in Fig.~\ref{fig: multiflow}(d). This time, the improvement due to the spatial-division rule is not as pronounced, since the spatial regions corresponding to `useful' repeater nodes for the two flows are not disjoint.

%In the case of multi-flow routing in a tree with our proposed repeater structure, it is possible to obtain an achievable upper bound on the rate region with global link-state knowledge. We consider multiple Alice-Bob pairs $\textrm{Alice}_i$ - $\textrm{Bob}_i$, each with the associated entanglement generation rate $R_i$. The rate region is defined by the set of equations:

%\begin{equation}
%\sum_{i \in \mathcal{I}} \frac{R_i}{q^{l_i-1}} \leq P_{1up}
%\end{equation}

%\noindent where $l_i$ is the length of the path between $\textrm{Alice}_i$-$\textrm{Bob}_i$ (in the case of a tree there is only one path between any two nodes) and $P_{1up}$ is the probability that atleast one pair $\textrm{Alice}_i$-$\textrm{Bob}_i$ in the index set $\mathcal{I}$ has a path between them. 

\section{Conclusions}

We proposed and analyzed quantum repeater protocols for entanglement generation in a quantum network in an architecture that uses the same elements as in linear repeater chains. We accounted for channel losses between repeater nodes and the probabilistic nature of entanglement swaps at each repeater stemming from device inefficiencies as well as the probabilistic nature of Bell-state measurements (e.g., due to inherent limitations of using linear optics and lossy detectors). The rate attained for a single entanglement-generation flow can far outperform that attainable over a linear repeater chain, even when the nodes only have local link-state knowledge, due to the multi-path routing advantage. We also proposed a modified version of our routing protocol for supporting simultaneous entanglement generation flows between multiple Alice-Bob pairs. We found multi-flow entanglement routing strategies that outperform the rate region attained when each repeater simply time shares among each flow. Our results suggest that building and connecting quantum repeaters in non-trivial network topologies could provide a substantial benefit over linear repeater chains alone. Seen another way, given constraints on the number and quality of quantum memories, link losses between nodes, and limited and imperfect processing capabilities at repeater nodes, a 2D network topology can outperform the repeater-less rate-vs.-distance upper limits~\cite{2014.NatureComm.Takeoka-Wilde.TGWBound, 2017.NatComm.Pirandola-Banchi.QuantCommUltRate} more easily than a linear repeater chain connecting the communicating parties. 

Our work has also opened a number of new questions. Even in our simplified model---an abstraction that applies when the only source of imperfection at each component (including the quantum memories) is pure loss---the rate-optimal protocol remains open. Since our protocol only requires a quantum memory to hold a qubit for one entanglement attempt between neighboring stations, photon loss would indeed be the major source of imperfection in many implementations of the protocol. Accounting for more general errors would require purification of entanglement~\cite{1996.PRL.Deutsch-Sanpera.PrivAmp, 1996.PRL.Bennett-Wootters.Purification, 1998.PRL.Briegel-Zoller.BDCZ}, i.e., converting several poorer-quality EPR pairs into a few good ones using local quantum operations and classical communication, accounting which will require us to introduce the Fidelity of shared entanglement at intermediate steps of the protocol. Furthermore, we restricted our analysis to the same operation used in the nodes of a linear repeater chain: 2-qubit measurements. Being able to perform multi-qubit unitary operations and multi-qubit measurements at repeater nodes (e.g., a 3-qubit GHZ projection across three locally held qubits) may improve the achievable rate regions. The idea of using a distance metric to choose the measurements at the repeater station could be used in protocols that use measurements of more than two qubits as well. Finally, it will be interesting to consider repeater protocols for the distillation of multi-partite entanglement shared between more than two parties, and a repeater network that can support multiple simultaneous flows of generation of multi-partite entanglement.

\begin{acknowledgments}
SG would like to thank Stefano Pirandola, Zachary Dutton, and Dongning Guo for valuable discussions. MP, DE and LJ acknowledge support from the Air Force Office of Scientific Research MURI (FA9550-14-1-0052) and the Army Research Laboratory (ARL) Center for Distributed Quantum Information (CDQI). SG, HK, PB, MP and DE would like to acknowledge the Office of Naval Research program {\em Communications and Networking with Quantum Operationally-Secure Technology for Maritime Deployment} (CONQUEST), awarded under prime contract number N00014-16-C-2069. SG, DT, PB and LT acknowledge the ARL DAIS-ITA program; this work would not have occurred without the collaborations and ideas seeded by this program.
\end{acknowledgments}

\bibliography{Qnetworkflowpaper}

%merlin.mbs apsrev4-1.bst 2010-07-25 4.21a (PWD, AO, DPC) hacked
%Control: key (0)
%Control: author (8) initials jnrlst
%Control: editor formatted (1) identically to author
%Control: production of article title (-1) disabled
%Control: page (0) single
%Control: year (1) truncated
%Control: production of eprint (0) enabled
\begin{thebibliography}{49}%
\makeatletter
\providecommand \@ifxundefined [1]{%
 \@ifx{#1\undefined}
}%
\providecommand \@ifnum [1]{%
 \ifnum #1\expandafter \@firstoftwo
 \else \expandafter \@secondoftwo
 \fi
}%
\providecommand \@ifx [1]{%
 \ifx #1\expandafter \@firstoftwo
 \else \expandafter \@secondoftwo
 \fi
}%
\providecommand \natexlab [1]{#1}%
\providecommand \enquote  [1]{``#1''}%
\providecommand \bibnamefont  [1]{#1}%
\providecommand \bibfnamefont [1]{#1}%
\providecommand \citenamefont [1]{#1}%
\providecommand \href@noop [0]{\@secondoftwo}%
\providecommand \href [0]{\begingroup \@sanitize@url \@href}%
\providecommand \@href[1]{\@@startlink{#1}\@@href}%
\providecommand \@@href[1]{\endgroup#1\@@endlink}%
\providecommand \@sanitize@url [0]{\catcode `\\12\catcode `\$12\catcode
  `\&12\catcode `\#12\catcode `\^12\catcode `\_12\catcode `\%12\relax}%
\providecommand \@@startlink[1]{}%
\providecommand \@@endlink[0]{}%
\providecommand \url  [0]{\begingroup\@sanitize@url \@url }%
\providecommand \@url [1]{\endgroup\@href {#1}{\urlprefix }}%
\providecommand \urlprefix  [0]{URL }%
\providecommand \Eprint [0]{\href }%
\providecommand \doibase [0]{http://dx.doi.org/}%
\providecommand \selectlanguage [0]{\@gobble}%
\providecommand \bibinfo  [0]{\@secondoftwo}%
\providecommand \bibfield  [0]{\@secondoftwo}%
\providecommand \translation [1]{[#1]}%
\providecommand \BibitemOpen [0]{}%
\providecommand \bibitemStop [0]{}%
\providecommand \bibitemNoStop [0]{.\EOS\space}%
\providecommand \EOS [0]{\spacefactor3000\relax}%
\providecommand \BibitemShut  [1]{\csname bibitem#1\endcsname}%
\let\auto@bib@innerbib\@empty
%</preamble>
\bibitem [{\citenamefont {Kimble}(2008)}]{2008.Nature.Kimble.QuantumInternet}%
  \BibitemOpen
  \bibfield  {author} {\bibinfo {author} {\bibfnamefont {H.~J.}\ \bibnamefont
  {Kimble}},\ }\href {\doibase 10.1038/nature07127} {\bibfield  {journal}
  {\bibinfo  {journal} {Nature}\ }\textbf {\bibinfo {volume} {453}},\ \bibinfo
  {pages} {1023} (\bibinfo {year} {2008})}\BibitemShut {NoStop}%
\bibitem [{\citenamefont {Ekert}(1991)}]{1991.PRL.Ekert.QCrypt}%
  \BibitemOpen
  \bibfield  {author} {\bibinfo {author} {\bibfnamefont {A.~K.}\ \bibnamefont
  {Ekert}},\ }\href {\doibase 10.1103/PhysRevLett.67.661} {\bibfield  {journal}
  {\bibinfo  {journal} {Physical Review Letters}\ }\textbf {\bibinfo {volume}
  {67}},\ \bibinfo {pages} {661} (\bibinfo {year} {1991})}\BibitemShut
  {NoStop}%
\bibitem [{\citenamefont {Bennett}\ and\ \citenamefont
  {Brassard}(2014)}]{2014.TheoCompSc.Bennett-Brassard.BB84}%
  \BibitemOpen
  \bibfield  {author} {\bibinfo {author} {\bibfnamefont {C.~H.}\ \bibnamefont
  {Bennett}}\ and\ \bibinfo {author} {\bibfnamefont {G.}~\bibnamefont
  {Brassard}},\ }\href {\doibase 10.1016/j.tcs.2014.05.025} {\bibfield
  {journal} {\bibinfo  {journal} {Theoretical Computer Science}\ }\textbf
  {\bibinfo {volume} {560}},\ \bibinfo {pages} {7} (\bibinfo {year}
  {2014})}\BibitemShut {NoStop}%
\bibitem [{\citenamefont {Cirac}\ \emph {et~al.}(1999)\citenamefont {Cirac},
  \citenamefont {Ekert}, \citenamefont {Huelga},\ and\ \citenamefont
  {Macchiavello}}]{1999.PRA.Cirac-Macchiavello.DistQC}%
  \BibitemOpen
  \bibfield  {author} {\bibinfo {author} {\bibfnamefont {J.}~\bibnamefont
  {Cirac}}, \bibinfo {author} {\bibfnamefont {A.}~\bibnamefont {Ekert}},
  \bibinfo {author} {\bibfnamefont {S.}~\bibnamefont {Huelga}}, \ and\ \bibinfo
  {author} {\bibfnamefont {C.}~\bibnamefont {Macchiavello}},\ }\href
  {https://journals.aps.org/pra/abstract/10.1103/PhysRevA.59.4249} {\bibfield
  {journal} {\bibinfo  {journal} {Physical Review A}\ } (\bibinfo {year}
  {1999})}\BibitemShut {NoStop}%
\bibitem [{\citenamefont {Gottesman}\ \emph {et~al.}(2012)\citenamefont
  {Gottesman}, \citenamefont {Jennewein},\ and\ \citenamefont
  {Croke}}]{2012.PRL.Gottesman-Croke.QRepTelescope}%
  \BibitemOpen
  \bibfield  {author} {\bibinfo {author} {\bibfnamefont {D.}~\bibnamefont
  {Gottesman}}, \bibinfo {author} {\bibfnamefont {T.}~\bibnamefont
  {Jennewein}}, \ and\ \bibinfo {author} {\bibfnamefont {S.}~\bibnamefont
  {Croke}},\ }\href {\doibase 10.1103/PhysRevLett.109.070503} {\bibfield
  {journal} {\bibinfo  {journal} {Physical Review Letters}\ }\textbf {\bibinfo
  {volume} {109}},\ \bibinfo {pages} {070503} (\bibinfo {year}
  {2012})}\BibitemShut {NoStop}%
\bibitem [{\citenamefont {K{\'{o}}m{\'{a}}r}\ \emph {et~al.}(2014)\citenamefont
  {K{\'{o}}m{\'{a}}r}, \citenamefont {Kessler}, \citenamefont {Bishof},
  \citenamefont {Jiang}, \citenamefont {S{\o}rensen}, \citenamefont {Ye},\ and\
  \citenamefont {Lukin}}]{2014.NatPhys.Komar-Lukin.QuantClocks}%
  \BibitemOpen
  \bibfield  {author} {\bibinfo {author} {\bibfnamefont {P.}~\bibnamefont
  {K{\'{o}}m{\'{a}}r}}, \bibinfo {author} {\bibfnamefont {E.~M.}\ \bibnamefont
  {Kessler}}, \bibinfo {author} {\bibfnamefont {M.}~\bibnamefont {Bishof}},
  \bibinfo {author} {\bibfnamefont {L.}~\bibnamefont {Jiang}}, \bibinfo
  {author} {\bibfnamefont {A.~S.}\ \bibnamefont {S{\o}rensen}}, \bibinfo
  {author} {\bibfnamefont {J.}~\bibnamefont {Ye}}, \ and\ \bibinfo {author}
  {\bibfnamefont {M.~D.}\ \bibnamefont {Lukin}},\ }\href {\doibase
  10.1038/nphys3000} {\bibfield  {journal} {\bibinfo  {journal} {Nature
  Physics}\ }\textbf {\bibinfo {volume} {10}},\ \bibinfo {pages} {582}
  (\bibinfo {year} {2014})},\ \Eprint {http://arxiv.org/abs/arXiv:1310.6045v1}
  {arXiv:arXiv:1310.6045v1} \BibitemShut {NoStop}%
\bibitem [{\citenamefont {Broadbent}\ \emph {et~al.}(2009)\citenamefont
  {Broadbent}, \citenamefont {Fitzsimons},\ and\ \citenamefont
  {Kashefi}}]{2009.FOCS.Broadbent-Kashefi.BlindQC}%
  \BibitemOpen
  \bibfield  {author} {\bibinfo {author} {\bibfnamefont {A.}~\bibnamefont
  {Broadbent}}, \bibinfo {author} {\bibfnamefont {J.}~\bibnamefont
  {Fitzsimons}}, \ and\ \bibinfo {author} {\bibfnamefont {E.}~\bibnamefont
  {Kashefi}},\ }in\ \href {\doibase 10.1109/FOCS.2009.36} {\emph {\bibinfo
  {booktitle} {2009 50th Annual IEEE Symposium on Foundations of Computer
  Science}}}\ (\bibinfo  {publisher} {IEEE},\ \bibinfo {year} {2009})\ pp.\
  \bibinfo {pages} {517--526}\BibitemShut {NoStop}%
\bibitem [{\citenamefont {Guha}\ \emph {et~al.}(2008)\citenamefont {Guha},
  \citenamefont {Hogg}, \citenamefont {Fattal}, \citenamefont {Spiller},\ and\
  \citenamefont {Beausoleil}}]{2008.IJQI.Guha-Beausoleil.QAuction}%
  \BibitemOpen
  \bibfield  {author} {\bibinfo {author} {\bibfnamefont {S.}~\bibnamefont
  {Guha}}, \bibinfo {author} {\bibfnamefont {T.}~\bibnamefont {Hogg}}, \bibinfo
  {author} {\bibfnamefont {D.}~\bibnamefont {Fattal}}, \bibinfo {author}
  {\bibfnamefont {T.}~\bibnamefont {Spiller}}, \ and\ \bibinfo {author}
  {\bibfnamefont {R.~G.}\ \bibnamefont {Beausoleil}},\ }\href {\doibase
  10.1142/S0219749908004183} {\bibfield  {journal} {\bibinfo  {journal}
  {International Journal of Quantum Information}\ }\textbf {\bibinfo {volume}
  {06}},\ \bibinfo {pages} {815} (\bibinfo {year} {2008})},\ \Eprint
  {http://arxiv.org/abs/arXiv:0707.2051v1} {arXiv:arXiv:0707.2051v1}
  \BibitemShut {NoStop}%
\bibitem [{\citenamefont {Hensen}\ \emph {et~al.}(2015)\citenamefont {Hensen},
  \citenamefont {Bernien}, \citenamefont {Dr{\'{e}}au}, \citenamefont
  {Reiserer}, \citenamefont {Kalb}, \citenamefont {Blok}, \citenamefont
  {Ruitenberg}, \citenamefont {Vermeulen}, \citenamefont {Schouten},
  \citenamefont {Abell{\'{a}}n}, \citenamefont {Amaya}, \citenamefont
  {Pruneri}, \citenamefont {Mitchell}, \citenamefont {Markham}, \citenamefont
  {Twitchen}, \citenamefont {Elkouss}, \citenamefont {Wehner}, \citenamefont
  {Taminiau},\ and\ \citenamefont
  {Hanson}}]{2015.Nature.Hensen-Hanson.BellTest}%
  \BibitemOpen
  \bibfield  {author} {\bibinfo {author} {\bibfnamefont {B.}~\bibnamefont
  {Hensen}}, \bibinfo {author} {\bibfnamefont {H.}~\bibnamefont {Bernien}},
  \bibinfo {author} {\bibfnamefont {A.~E.}\ \bibnamefont {Dr{\'{e}}au}},
  \bibinfo {author} {\bibfnamefont {A.}~\bibnamefont {Reiserer}}, \bibinfo
  {author} {\bibfnamefont {N.}~\bibnamefont {Kalb}}, \bibinfo {author}
  {\bibfnamefont {M.~S.}\ \bibnamefont {Blok}}, \bibinfo {author}
  {\bibfnamefont {J.}~\bibnamefont {Ruitenberg}}, \bibinfo {author}
  {\bibfnamefont {R.~F.~L.}\ \bibnamefont {Vermeulen}}, \bibinfo {author}
  {\bibfnamefont {R.~N.}\ \bibnamefont {Schouten}}, \bibinfo {author}
  {\bibfnamefont {C.}~\bibnamefont {Abell{\'{a}}n}}, \bibinfo {author}
  {\bibfnamefont {W.}~\bibnamefont {Amaya}}, \bibinfo {author} {\bibfnamefont
  {V.}~\bibnamefont {Pruneri}}, \bibinfo {author} {\bibfnamefont {M.~W.}\
  \bibnamefont {Mitchell}}, \bibinfo {author} {\bibfnamefont {M.}~\bibnamefont
  {Markham}}, \bibinfo {author} {\bibfnamefont {D.~J.}\ \bibnamefont
  {Twitchen}}, \bibinfo {author} {\bibfnamefont {D.}~\bibnamefont {Elkouss}},
  \bibinfo {author} {\bibfnamefont {S.}~\bibnamefont {Wehner}}, \bibinfo
  {author} {\bibfnamefont {T.~H.}\ \bibnamefont {Taminiau}}, \ and\ \bibinfo
  {author} {\bibfnamefont {R.}~\bibnamefont {Hanson}},\ }\href {\doibase
  10.1038/nature15759} {\bibfield  {journal} {\bibinfo  {journal} {Nature}\
  }\textbf {\bibinfo {volume} {526}},\ \bibinfo {pages} {682} (\bibinfo {year}
  {2015})},\ \Eprint {http://arxiv.org/abs/1508.05949} {arXiv:1508.05949}
  \BibitemShut {NoStop}%
\bibitem [{\citenamefont {Takeoka}\ \emph {et~al.}(2014)\citenamefont
  {Takeoka}, \citenamefont {Guha},\ and\ \citenamefont
  {Wilde}}]{2014.NatureComm.Takeoka-Wilde.TGWBound}%
  \BibitemOpen
  \bibfield  {author} {\bibinfo {author} {\bibfnamefont {M.}~\bibnamefont
  {Takeoka}}, \bibinfo {author} {\bibfnamefont {S.}~\bibnamefont {Guha}}, \
  and\ \bibinfo {author} {\bibfnamefont {M.~M.}\ \bibnamefont {Wilde}},\ }\href
  {\doibase 10.1038/ncomms6235} {\bibfield  {journal} {\bibinfo  {journal}
  {Nature communications}\ }\textbf {\bibinfo {volume} {5}},\ \bibinfo {pages}
  {5235} (\bibinfo {year} {2014})}\BibitemShut {NoStop}%
\bibitem [{\citenamefont {Pirandola}\ \emph {et~al.}(2017)\citenamefont
  {Pirandola}, \citenamefont {Laurenza}, \citenamefont {Ottaviani},\ and\
  \citenamefont {Banchi}}]{2017.NatComm.Pirandola-Banchi.QuantCommUltRate}%
  \BibitemOpen
  \bibfield  {author} {\bibinfo {author} {\bibfnamefont {S.}~\bibnamefont
  {Pirandola}}, \bibinfo {author} {\bibfnamefont {R.}~\bibnamefont {Laurenza}},
  \bibinfo {author} {\bibfnamefont {C.}~\bibnamefont {Ottaviani}}, \ and\
  \bibinfo {author} {\bibfnamefont {L.}~\bibnamefont {Banchi}},\ }\href
  {\doibase 10.1038/ncomms15043} {\bibfield  {journal} {\bibinfo  {journal}
  {Nature Communications}\ }\textbf {\bibinfo {volume} {8}},\ \bibinfo {pages}
  {15043} (\bibinfo {year} {2017})},\ \Eprint {http://arxiv.org/abs/1510.08863}
  {arXiv:1510.08863} \BibitemShut {NoStop}%
\bibitem [{\citenamefont {Vahdat}\ and\ \citenamefont
  {Becker}(2000)}]{2000.TechRep.Vahdat-Becker.EpRoutAdhoc}%
  \BibitemOpen
  \bibfield  {author} {\bibinfo {author} {\bibfnamefont {A.}~\bibnamefont
  {Vahdat}}\ and\ \bibinfo {author} {\bibfnamefont {D.}~\bibnamefont
  {Becker}},\ }\href {\doibase 10.1.1.34.6151} {\bibfield  {journal} {\bibinfo
  {journal} {Technical report number CS-200006, Duke University}\ ,\ \bibinfo
  {pages} {1}} (\bibinfo {year} {2000})},\ \Eprint
  {http://arxiv.org/abs/1001.3405} {arXiv:1001.3405} \BibitemShut {NoStop}%
\bibitem [{\citenamefont {Biswas}\ and\ \citenamefont
  {Morris}(2005)}]{2005.Sigcomm.Biswas-Morrix.ExOR}%
  \BibitemOpen
  \bibfield  {author} {\bibinfo {author} {\bibfnamefont {S.}~\bibnamefont
  {Biswas}}\ and\ \bibinfo {author} {\bibfnamefont {R.}~\bibnamefont
  {Morris}},\ }\href {\doibase 10.1145/1090191.1080108} {\bibfield  {journal}
  {\bibinfo  {journal} {Sigcomm}\ } (\bibinfo {year} {2005}),\
  10.1145/1090191.1080108}\BibitemShut {NoStop}%
\bibitem [{\citenamefont {Wootters}\ and\ \citenamefont
  {Zurek}(1982)}]{1982.Nature.Wootters-Zurek.NoCloning}%
  \BibitemOpen
  \bibfield  {author} {\bibinfo {author} {\bibfnamefont {W.~K.}\ \bibnamefont
  {Wootters}}\ and\ \bibinfo {author} {\bibfnamefont {W.~H.}\ \bibnamefont
  {Zurek}},\ }\href {\doibase 10.1038/299802a0} {\bibfield  {journal} {\bibinfo
   {journal} {Nature}\ }\textbf {\bibinfo {volume} {299}},\ \bibinfo {pages}
  {802} (\bibinfo {year} {1982})}\BibitemShut {NoStop}%
\bibitem [{\citenamefont {Dieks}(1982)}]{1982.PLA.Dieks.EPRComm}%
  \BibitemOpen
  \bibfield  {author} {\bibinfo {author} {\bibfnamefont {D.}~\bibnamefont
  {Dieks}},\ }\href {\doibase 10.1016/0375-9601(82)90084-6} {\bibfield
  {journal} {\bibinfo  {journal} {Physics Letters A}\ }\textbf {\bibinfo
  {volume} {92}},\ \bibinfo {pages} {271} (\bibinfo {year} {1982})}\BibitemShut
  {NoStop}%
\bibitem [{\citenamefont {Jain}\ \emph {et~al.}(2004)\citenamefont {Jain},
  \citenamefont {Fall},\ and\ \citenamefont
  {Patra}}]{2004.SigComm.Jain-Patra.DTNRout}%
  \BibitemOpen
  \bibfield  {author} {\bibinfo {author} {\bibfnamefont {S.}~\bibnamefont
  {Jain}}, \bibinfo {author} {\bibfnamefont {K.}~\bibnamefont {Fall}}, \ and\
  \bibinfo {author} {\bibfnamefont {R.}~\bibnamefont {Patra}},\ }in\ \href
  {\doibase 10.1145/1015467.1015484} {\emph {\bibinfo {booktitle} {Proceedings
  of the 2004 conference on Applications, technologies, architectures, and
  protocols for computer communications - SIGCOMM '04}}},\ Vol.~\bibinfo
  {volume} {34}\ (\bibinfo  {publisher} {ACM Press},\ \bibinfo {address} {New
  York, New York, USA},\ \bibinfo {year} {2004})\ p.\ \bibinfo {pages}
  {145}\BibitemShut {NoStop}%
\bibitem [{\citenamefont {Burgess}\ \emph {et~al.}(2006)\citenamefont
  {Burgess}, \citenamefont {Gallagher}, \citenamefont {Jensen},\ and\
  \citenamefont {Levine}}]{2006.InfoComm.Burgess-Levine.MaxProp}%
  \BibitemOpen
  \bibfield  {author} {\bibinfo {author} {\bibfnamefont {J.}~\bibnamefont
  {Burgess}}, \bibinfo {author} {\bibfnamefont {B.}~\bibnamefont {Gallagher}},
  \bibinfo {author} {\bibfnamefont {D.}~\bibnamefont {Jensen}}, \ and\ \bibinfo
  {author} {\bibfnamefont {B.~N.}\ \bibnamefont {Levine}},\ }\href {\doibase
  10.1109/INFOCOM.2006.228} {\bibfield  {journal} {\bibinfo  {journal}
  {Proceedings - IEEE INFOCOM}\ }\textbf {\bibinfo {volume} {00}} (\bibinfo
  {year} {2006}),\ 10.1109/INFOCOM.2006.228}\BibitemShut {NoStop}%
\bibitem [{\citenamefont {Yuan}\ \emph {et~al.}(2008)\citenamefont {Yuan},
  \citenamefont {Chen}, \citenamefont {Zhao}, \citenamefont {Chen},
  \citenamefont {Schmiedmayer},\ and\ \citenamefont
  {Pan}}]{2008.Nature.Yuan-Pan.BDCZExp}%
  \BibitemOpen
  \bibfield  {author} {\bibinfo {author} {\bibfnamefont {Z.-S.}\ \bibnamefont
  {Yuan}}, \bibinfo {author} {\bibfnamefont {Y.-A.}\ \bibnamefont {Chen}},
  \bibinfo {author} {\bibfnamefont {B.}~\bibnamefont {Zhao}}, \bibinfo {author}
  {\bibfnamefont {S.}~\bibnamefont {Chen}}, \bibinfo {author} {\bibfnamefont
  {J.}~\bibnamefont {Schmiedmayer}}, \ and\ \bibinfo {author} {\bibfnamefont
  {J.-W.}\ \bibnamefont {Pan}},\ }\href {\doibase 10.1038/nature07241}
  {\bibfield  {journal} {\bibinfo  {journal} {Nature}\ }\textbf {\bibinfo
  {volume} {454}},\ \bibinfo {pages} {1098} (\bibinfo {year}
  {2008})}\BibitemShut {NoStop}%
\bibitem [{\citenamefont {Bernien}\ \emph {et~al.}(2013)\citenamefont
  {Bernien}, \citenamefont {Hensen}, \citenamefont {Pfaff}, \citenamefont
  {Koolstra}, \citenamefont {Blok}, \citenamefont {Robledo}, \citenamefont
  {Taminiau}, \citenamefont {Markham}, \citenamefont {Twitchen}, \citenamefont
  {Childress},\ and\ \citenamefont
  {Hanson}}]{2013.Nature.Bernien-Hanson.3ment}%
  \BibitemOpen
  \bibfield  {author} {\bibinfo {author} {\bibfnamefont {H.}~\bibnamefont
  {Bernien}}, \bibinfo {author} {\bibfnamefont {B.}~\bibnamefont {Hensen}},
  \bibinfo {author} {\bibfnamefont {W.}~\bibnamefont {Pfaff}}, \bibinfo
  {author} {\bibfnamefont {G.}~\bibnamefont {Koolstra}}, \bibinfo {author}
  {\bibfnamefont {M.~S.}\ \bibnamefont {Blok}}, \bibinfo {author}
  {\bibfnamefont {L.}~\bibnamefont {Robledo}}, \bibinfo {author} {\bibfnamefont
  {T.~H.}\ \bibnamefont {Taminiau}}, \bibinfo {author} {\bibfnamefont
  {M.}~\bibnamefont {Markham}}, \bibinfo {author} {\bibfnamefont {D.~J.}\
  \bibnamefont {Twitchen}}, \bibinfo {author} {\bibfnamefont {L.}~\bibnamefont
  {Childress}}, \ and\ \bibinfo {author} {\bibfnamefont {R.}~\bibnamefont
  {Hanson}},\ }\href {\doibase 10.1038/nature12016} {\bibfield  {journal}
  {\bibinfo  {journal} {Nature}\ }\textbf {\bibinfo {volume} {497}},\ \bibinfo
  {pages} {86} (\bibinfo {year} {2013})},\ \Eprint
  {http://arxiv.org/abs/1212.6136} {arXiv:1212.6136} \BibitemShut {NoStop}%
\bibitem [{\citenamefont {Olmschenk}\ \emph {et~al.}(2009)\citenamefont
  {Olmschenk}, \citenamefont {Matsukevich}, \citenamefont {Maunz},
  \citenamefont {Hayes}, \citenamefont {Duan},\ and\ \citenamefont
  {Monroe}}]{2009.Science.Olmschenk-Monroe.QuantTelMatQubit}%
  \BibitemOpen
  \bibfield  {author} {\bibinfo {author} {\bibfnamefont {S.}~\bibnamefont
  {Olmschenk}}, \bibinfo {author} {\bibfnamefont {D.~N.}\ \bibnamefont
  {Matsukevich}}, \bibinfo {author} {\bibfnamefont {P.}~\bibnamefont {Maunz}},
  \bibinfo {author} {\bibfnamefont {D.}~\bibnamefont {Hayes}}, \bibinfo
  {author} {\bibfnamefont {L.-M.}\ \bibnamefont {Duan}}, \ and\ \bibinfo
  {author} {\bibfnamefont {C.}~\bibnamefont {Monroe}},\ }\href {\doibase
  10.1126/science.1167209} {\bibfield  {journal} {\bibinfo  {journal}
  {Science}\ }\textbf {\bibinfo {volume} {323}},\ \bibinfo {pages} {486}
  (\bibinfo {year} {2009})},\ \Eprint {http://arxiv.org/abs/0907.5240}
  {arXiv:0907.5240} \BibitemShut {NoStop}%
\bibitem [{\citenamefont {Pan}\ \emph {et~al.}(1998)\citenamefont {Pan},
  \citenamefont {Bouwmeester}, \citenamefont {Weinfurter},\ and\ \citenamefont
  {Zeilinger}}]{1998.PRL.Pan-Zeilinger.EntSwapPhot}%
  \BibitemOpen
  \bibfield  {author} {\bibinfo {author} {\bibfnamefont {J.-W.}\ \bibnamefont
  {Pan}}, \bibinfo {author} {\bibfnamefont {D.}~\bibnamefont {Bouwmeester}},
  \bibinfo {author} {\bibfnamefont {H.}~\bibnamefont {Weinfurter}}, \ and\
  \bibinfo {author} {\bibfnamefont {A.}~\bibnamefont {Zeilinger}},\ }\href
  {\doibase 10.1103/PhysRevLett.80.3891} {\bibfield  {journal} {\bibinfo
  {journal} {Physical Review Letters}\ }\textbf {\bibinfo {volume} {80}},\
  \bibinfo {pages} {3891} (\bibinfo {year} {1998})}\BibitemShut {NoStop}%
\bibitem [{\citenamefont {Chou}\ \emph {et~al.}(2005)\citenamefont {Chou},
  \citenamefont {de~Riedmatten}, \citenamefont {Felinto}, \citenamefont
  {Polyakov}, \citenamefont {van Enk},\ and\ \citenamefont
  {Kimble}}]{2005.Nature.Chou-Kimble.MeasIndEnt}%
  \BibitemOpen
  \bibfield  {author} {\bibinfo {author} {\bibfnamefont {C.~W.}\ \bibnamefont
  {Chou}}, \bibinfo {author} {\bibfnamefont {H.}~\bibnamefont {de~Riedmatten}},
  \bibinfo {author} {\bibfnamefont {D.}~\bibnamefont {Felinto}}, \bibinfo
  {author} {\bibfnamefont {S.~V.}\ \bibnamefont {Polyakov}}, \bibinfo {author}
  {\bibfnamefont {S.~J.}\ \bibnamefont {van Enk}}, \ and\ \bibinfo {author}
  {\bibfnamefont {H.~J.}\ \bibnamefont {Kimble}},\ }\href {\doibase
  10.1038/nature04353} {\bibfield  {journal} {\bibinfo  {journal} {Nature}\
  }\textbf {\bibinfo {volume} {438}},\ \bibinfo {pages} {828} (\bibinfo {year}
  {2005})}\BibitemShut {NoStop}%
\bibitem [{\citenamefont {Moehring}\ \emph {et~al.}(2007)\citenamefont
  {Moehring}, \citenamefont {Maunz}, \citenamefont {Olmschenk}, \citenamefont
  {Younge}, \citenamefont {Matsukevich}, \citenamefont {Duan},\ and\
  \citenamefont {Monroe}}]{2007.Nature.EntSingleAtomBits}%
  \BibitemOpen
  \bibfield  {author} {\bibinfo {author} {\bibfnamefont {D.~L.}\ \bibnamefont
  {Moehring}}, \bibinfo {author} {\bibfnamefont {P.}~\bibnamefont {Maunz}},
  \bibinfo {author} {\bibfnamefont {S.}~\bibnamefont {Olmschenk}}, \bibinfo
  {author} {\bibfnamefont {K.~C.}\ \bibnamefont {Younge}}, \bibinfo {author}
  {\bibfnamefont {D.~N.}\ \bibnamefont {Matsukevich}}, \bibinfo {author}
  {\bibfnamefont {L.-M.}\ \bibnamefont {Duan}}, \ and\ \bibinfo {author}
  {\bibfnamefont {C.}~\bibnamefont {Monroe}},\ }\href {\doibase
  10.1038/nature06118} {\bibfield  {journal} {\bibinfo  {journal} {Nature}\
  }\textbf {\bibinfo {volume} {449}},\ \bibinfo {pages} {68} (\bibinfo {year}
  {2007})}\BibitemShut {NoStop}%
\bibitem [{Note1()}]{Note1}%
  \BibitemOpen
  \bibinfo {note} {Pirandola recently showed~\cite
  {2016.ArXiv.Pirandola.CapRepQC}, for an information-theoretic description of
  repeaters that are ideal fully-error-corrected universal quantum processors,
  that the optimal rate attainable for multi-path entanglement routing using
  such ideal repeaters is superior to the rate of a linear chain of ideal
  repeaters.}\BibitemShut {Stop}%
\bibitem [{\citenamefont {Guha}\ \emph {et~al.}(2015)\citenamefont {Guha},
  \citenamefont {Krovi}, \citenamefont {Fuchs}, \citenamefont {Dutton},
  \citenamefont {Slater}, \citenamefont {Simon},\ and\ \citenamefont
  {Tittel}}]{2015.PRA.Guha-Tittel.QRRateLossAnalysis}%
  \BibitemOpen
  \bibfield  {author} {\bibinfo {author} {\bibfnamefont {S.}~\bibnamefont
  {Guha}}, \bibinfo {author} {\bibfnamefont {H.}~\bibnamefont {Krovi}},
  \bibinfo {author} {\bibfnamefont {C.~A.}\ \bibnamefont {Fuchs}}, \bibinfo
  {author} {\bibfnamefont {Z.}~\bibnamefont {Dutton}}, \bibinfo {author}
  {\bibfnamefont {J.~A.}\ \bibnamefont {Slater}}, \bibinfo {author}
  {\bibfnamefont {C.}~\bibnamefont {Simon}}, \ and\ \bibinfo {author}
  {\bibfnamefont {W.}~\bibnamefont {Tittel}},\ }\href {\doibase
  10.1103/PhysRevA.92.022357} {\bibfield  {journal} {\bibinfo  {journal}
  {Physical Review A}\ }\textbf {\bibinfo {volume} {92}},\ \bibinfo {pages}
  {022357} (\bibinfo {year} {2015})}\BibitemShut {NoStop}%
\bibitem [{\citenamefont {Pant}\ \emph {et~al.}(2017)\citenamefont {Pant},
  \citenamefont {Krovi}, \citenamefont {Englund},\ and\ \citenamefont
  {Guha}}]{2017.PRA.Pant-Guha.AllOptRepRescources}%
  \BibitemOpen
  \bibfield  {author} {\bibinfo {author} {\bibfnamefont {M.}~\bibnamefont
  {Pant}}, \bibinfo {author} {\bibfnamefont {H.}~\bibnamefont {Krovi}},
  \bibinfo {author} {\bibfnamefont {D.}~\bibnamefont {Englund}}, \ and\
  \bibinfo {author} {\bibfnamefont {S.}~\bibnamefont {Guha}},\ }\href {\doibase
  10.1103/PhysRevA.95.012304} {\bibfield  {journal} {\bibinfo  {journal}
  {Physical Review A}\ }\textbf {\bibinfo {volume} {95}},\ \bibinfo {pages}
  {012304} (\bibinfo {year} {2017})},\ \Eprint
  {http://arxiv.org/abs/1603.01353} {arXiv:1603.01353} \BibitemShut {NoStop}%
\bibitem [{Note2()}]{Note2}%
  \BibitemOpen
  \bibinfo {note} {Achievability of $\eta $ ebits/mode on a linear repeater
  chain---with $\eta $ being the transmissivity of the minimum-transmissivity
  link---with unrestricted processing at the repeater nodes follows from the
  single-photon dual-rail encoded realization of the Ekert-91 protocol~\cite
  {1991.PRL.Ekert.QCrypt} together with ideal BSMs, thereby establishing that
  the rate achieved by a linear chain of ideal repeaters is superior to that
  attained without any repeater assistance. Azuma {\protect \em et al.}
  generalized the TGW bound to an upper bound on the rate achievable for a
  single entanglement generation flow (line repeater being a special case),
  which established $\protect \qopname \relax o{log}_2[(1+\eta )/(1-\eta )]
  \approx 2.89\eta $ for $\eta \ll 1$ as an upper bound to the rate~\cite
  {2016.NatComm.Azuma-Lo.QNetrateloss}. Pirandola generalized the PLOB upper
  bound~\cite {2016.ArXiv.Pirandola.CapRepQC} to repeater networks and proved a
  matching lower bound, showing that with an information-theoretic description
  of quantum repeaters, i.e., that are ideal fully-error-corrected universal
  quantum processors, that the optimal rate attainable for entanglement
  generation on a linear repeater chain is given by $\protect \qopname \relax
  o{log}_2[1/(1-\eta )] \approx 1.44\eta $ for $\eta \ll 1$.}\BibitemShut
  {Stop}%
\bibitem [{\citenamefont {Deutsch}\ \emph {et~al.}(1996)\citenamefont
  {Deutsch}, \citenamefont {Ekert}, \citenamefont {Jozsa}, \citenamefont
  {Macchiavello}, \citenamefont {Popescu},\ and\ \citenamefont
  {Sanpera}}]{1996.PRL.Deutsch-Sanpera.PrivAmp}%
  \BibitemOpen
  \bibfield  {author} {\bibinfo {author} {\bibfnamefont {D.}~\bibnamefont
  {Deutsch}}, \bibinfo {author} {\bibfnamefont {A.}~\bibnamefont {Ekert}},
  \bibinfo {author} {\bibfnamefont {R.}~\bibnamefont {Jozsa}}, \bibinfo
  {author} {\bibfnamefont {C.}~\bibnamefont {Macchiavello}}, \bibinfo {author}
  {\bibfnamefont {S.}~\bibnamefont {Popescu}}, \ and\ \bibinfo {author}
  {\bibfnamefont {A.}~\bibnamefont {Sanpera}},\ }\href {\doibase
  10.1103/PhysRevLett.77.2818} {\bibfield  {journal} {\bibinfo  {journal}
  {Physical Review Letters}\ }\textbf {\bibinfo {volume} {77}},\ \bibinfo
  {pages} {2818} (\bibinfo {year} {1996})}\BibitemShut {NoStop}%
\bibitem [{\citenamefont {Bennett}\ \emph {et~al.}(1996)\citenamefont
  {Bennett}, \citenamefont {Brassard}, \citenamefont {Popescu}, \citenamefont
  {Schumacher}, \citenamefont {Smolin},\ and\ \citenamefont
  {Wootters}}]{1996.PRL.Bennett-Wootters.Purification}%
  \BibitemOpen
  \bibfield  {author} {\bibinfo {author} {\bibfnamefont {C.~H.}\ \bibnamefont
  {Bennett}}, \bibinfo {author} {\bibfnamefont {G.}~\bibnamefont {Brassard}},
  \bibinfo {author} {\bibfnamefont {S.}~\bibnamefont {Popescu}}, \bibinfo
  {author} {\bibfnamefont {B.}~\bibnamefont {Schumacher}}, \bibinfo {author}
  {\bibfnamefont {J.~A.}\ \bibnamefont {Smolin}}, \ and\ \bibinfo {author}
  {\bibfnamefont {W.~K.}\ \bibnamefont {Wootters}},\ }\href {\doibase
  10.1103/PhysRevLett.76.722} {\bibfield  {journal} {\bibinfo  {journal}
  {Physical Review Letters}\ }\textbf {\bibinfo {volume} {76}},\ \bibinfo
  {pages} {722} (\bibinfo {year} {1996})}\BibitemShut {NoStop}%
\bibitem [{\citenamefont {Briegel}\ \emph {et~al.}(1998)\citenamefont
  {Briegel}, \citenamefont {D{\"{u}}r}, \citenamefont {Cirac},\ and\
  \citenamefont {Zoller}}]{1998.PRL.Briegel-Zoller.BDCZ}%
  \BibitemOpen
  \bibfield  {author} {\bibinfo {author} {\bibfnamefont {H.-J.}\ \bibnamefont
  {Briegel}}, \bibinfo {author} {\bibfnamefont {W.}~\bibnamefont {D{\"{u}}r}},
  \bibinfo {author} {\bibfnamefont {J.}~\bibnamefont {Cirac}}, \ and\ \bibinfo
  {author} {\bibfnamefont {P.}~\bibnamefont {Zoller}},\ }\href {\doibase
  10.1103/PhysRevLett.81.5932} {\bibfield  {journal} {\bibinfo  {journal}
  {Physical Review Letters}\ }\textbf {\bibinfo {volume} {81}},\ \bibinfo
  {pages} {5932} (\bibinfo {year} {1998})}\BibitemShut {NoStop}%
\bibitem [{\citenamefont {Jiang}\ \emph {et~al.}(2009)\citenamefont {Jiang},
  \citenamefont {Taylor}, \citenamefont {Nemoto}, \citenamefont {Munro},
  \citenamefont {{Van Meter}},\ and\ \citenamefont
  {Lukin}}]{2009.PRA.Jiang-Lukin.3genQR}%
  \BibitemOpen
  \bibfield  {author} {\bibinfo {author} {\bibfnamefont {L.}~\bibnamefont
  {Jiang}}, \bibinfo {author} {\bibfnamefont {J.~M.}\ \bibnamefont {Taylor}},
  \bibinfo {author} {\bibfnamefont {K.}~\bibnamefont {Nemoto}}, \bibinfo
  {author} {\bibfnamefont {W.~J.}\ \bibnamefont {Munro}}, \bibinfo {author}
  {\bibfnamefont {R.}~\bibnamefont {{Van Meter}}}, \ and\ \bibinfo {author}
  {\bibfnamefont {M.~D.}\ \bibnamefont {Lukin}},\ }\href {\doibase
  10.1103/PhysRevA.79.032325} {\bibfield  {journal} {\bibinfo  {journal}
  {Physical Review A}\ }\textbf {\bibinfo {volume} {79}},\ \bibinfo {pages}
  {032325} (\bibinfo {year} {2009})}\BibitemShut {NoStop}%
\bibitem [{Note3()}]{Note3}%
  \BibitemOpen
  \bibinfo {note} {The achievability of $-\protect \qopname \relax
  o{log}_2(1-\eta )$ ebits per mode of secret communication rate over the lossy
  channel (with two way authenticated public classical communication) was first
  proven in 2009 by Pirandola {\protect \em et al.}~\cite
  {2009.PRL.Pirandola-Lloyd.SKCLB}. In 2014, Takeoka {\protect \em et al.}
  proved an upper bound to the secret-key agreement capacity, $\protect
  \qopname \relax o{log}_2[(1+\eta )/(1-\eta )]$ ebits per mode~\cite
  {2014.NatureComm.Takeoka-Wilde.TGWBound}, which equals $\approx 2.88\eta $
  ebits per mode when $\eta \ll 1$, thereby establishing that the rate attained
  by {\protect \em any} protocol must decay linearly with the channel's
  transmissivity and hence exponentially with distance $L$ in optical fiber
  (since $\eta \sim e^{\alpha L}$). In 2015, Pirandola {\protect \em et al.}
  proved an improved (weak converse) upper bound of $-\protect \qopname \relax
  o{log}_2(1-\eta )$ ebits per mode, which established that as the secret key
  agreement capacity of the pure loss bosonic channel~\cite
  {2017.NatComm.Pirandola-Banchi.QuantCommUltRate}. In 2017, Wilde {\protect
  \em et al.} proved $-\protect \qopname \relax o{log}_2(1-\eta )$ ebits per
  mode as a strong converse upper bound to the secret-key agreement
  capacity~\cite
  {2017.IEEETransInfTheo.Wilde-Berta.PricComQChanStrongConv}.}\BibitemShut
  {Stop}%
\bibitem [{\citenamefont {Muralidharan}\ \emph {et~al.}(2016)\citenamefont
  {Muralidharan}, \citenamefont {Li}, \citenamefont {Kim}, \citenamefont
  {L{\"{u}}tkenhaus}, \citenamefont {Lukin},\ and\ \citenamefont
  {Jiang}}]{2016.SciRep.Muralidharan-Jiang.RepeaterGen}%
  \BibitemOpen
  \bibfield  {author} {\bibinfo {author} {\bibfnamefont {S.}~\bibnamefont
  {Muralidharan}}, \bibinfo {author} {\bibfnamefont {L.}~\bibnamefont {Li}},
  \bibinfo {author} {\bibfnamefont {J.}~\bibnamefont {Kim}}, \bibinfo {author}
  {\bibfnamefont {N.}~\bibnamefont {L{\"{u}}tkenhaus}}, \bibinfo {author}
  {\bibfnamefont {M.~D.}\ \bibnamefont {Lukin}}, \ and\ \bibinfo {author}
  {\bibfnamefont {L.}~\bibnamefont {Jiang}},\ }\href {\doibase
  10.1038/srep20463} {\bibfield  {journal} {\bibinfo  {journal} {Scientific
  Reports}\ }\textbf {\bibinfo {volume} {6}},\ \bibinfo {pages} {20463}
  (\bibinfo {year} {2016})},\ \Eprint {http://arxiv.org/abs/1509.08435}
  {arXiv:1509.08435} \BibitemShut {NoStop}%
\bibitem [{\citenamefont {Sinclair}\ \emph {et~al.}(2014)\citenamefont
  {Sinclair}, \citenamefont {Saglamyurek}, \citenamefont {Mallahzadeh},
  \citenamefont {Slater}, \citenamefont {George}, \citenamefont {Ricken},
  \citenamefont {Hedges}, \citenamefont {Oblak}, \citenamefont {Simon},
  \citenamefont {Sohler},\ and\ \citenamefont
  {Tittel}}]{2014.PRL.Sinclair-Tittel.AtomFreqCombRep}%
  \BibitemOpen
  \bibfield  {author} {\bibinfo {author} {\bibfnamefont {N.}~\bibnamefont
  {Sinclair}}, \bibinfo {author} {\bibfnamefont {E.}~\bibnamefont
  {Saglamyurek}}, \bibinfo {author} {\bibfnamefont {H.}~\bibnamefont
  {Mallahzadeh}}, \bibinfo {author} {\bibfnamefont {J.~A.}\ \bibnamefont
  {Slater}}, \bibinfo {author} {\bibfnamefont {M.}~\bibnamefont {George}},
  \bibinfo {author} {\bibfnamefont {R.}~\bibnamefont {Ricken}}, \bibinfo
  {author} {\bibfnamefont {M.~P.}\ \bibnamefont {Hedges}}, \bibinfo {author}
  {\bibfnamefont {D.}~\bibnamefont {Oblak}}, \bibinfo {author} {\bibfnamefont
  {C.}~\bibnamefont {Simon}}, \bibinfo {author} {\bibfnamefont
  {W.}~\bibnamefont {Sohler}}, \ and\ \bibinfo {author} {\bibfnamefont
  {W.}~\bibnamefont {Tittel}},\ }\href {\doibase
  10.1103/PhysRevLett.113.053603} {\bibfield  {journal} {\bibinfo  {journal}
  {Physical review letters}\ }\textbf {\bibinfo {volume} {113}},\ \bibinfo
  {pages} {053603} (\bibinfo {year} {2014})}\BibitemShut {NoStop}%
\bibitem [{\citenamefont {Azuma}\ \emph {et~al.}(2015)\citenamefont {Azuma},
  \citenamefont {Tamaki},\ and\ \citenamefont
  {Lo}}]{2015.NatureComm.Azuma-Lo.AllOptRep}%
  \BibitemOpen
  \bibfield  {author} {\bibinfo {author} {\bibfnamefont {K.}~\bibnamefont
  {Azuma}}, \bibinfo {author} {\bibfnamefont {K.}~\bibnamefont {Tamaki}}, \
  and\ \bibinfo {author} {\bibfnamefont {H.-K.}\ \bibnamefont {Lo}},\ }\href
  {\doibase 10.1038/ncomms7787} {\bibfield  {journal} {\bibinfo  {journal}
  {Nature communications}\ }\textbf {\bibinfo {volume} {6}},\ \bibinfo {pages}
  {6787} (\bibinfo {year} {2015})}\BibitemShut {NoStop}%
\bibitem [{\citenamefont {Ewert}\ \emph {et~al.}(2016)\citenamefont {Ewert},
  \citenamefont {Bergmann},\ and\ \citenamefont {van
  Loock}}]{2016.PRL.Ewert-VanLoock.QPCPhotRep}%
  \BibitemOpen
  \bibfield  {author} {\bibinfo {author} {\bibfnamefont {F.}~\bibnamefont
  {Ewert}}, \bibinfo {author} {\bibfnamefont {M.}~\bibnamefont {Bergmann}}, \
  and\ \bibinfo {author} {\bibfnamefont {P.}~\bibnamefont {van Loock}},\ }\href
  {\doibase 10.1103/PhysRevLett.117.210501} {\bibfield  {journal} {\bibinfo
  {journal} {Physical Review Letters}\ }\textbf {\bibinfo {volume} {117}},\
  \bibinfo {pages} {210501} (\bibinfo {year} {2016})}\BibitemShut {NoStop}%
\bibitem [{\citenamefont {Pirandola}(2016)}]{2016.ArXiv.Pirandola.CapRepQC}%
  \BibitemOpen
  \bibfield  {author} {\bibinfo {author} {\bibfnamefont {S.}~\bibnamefont
  {Pirandola}},\ }\href {http://arxiv.org/abs/1601.00966} {\bibfield  {journal}
  {\bibinfo  {journal} {arXiv preprint}\ } (\bibinfo {year} {2016})},\ \Eprint
  {http://arxiv.org/abs/1601.00966} {arXiv:1601.00966} \BibitemShut {NoStop}%
\bibitem [{\citenamefont {Azuma}\ \emph {et~al.}(2016)\citenamefont {Azuma},
  \citenamefont {Mizutani},\ and\ \citenamefont
  {Lo}}]{2016.NatComm.Azuma-Lo.QNetrateloss}%
  \BibitemOpen
  \bibfield  {author} {\bibinfo {author} {\bibfnamefont {K.}~\bibnamefont
  {Azuma}}, \bibinfo {author} {\bibfnamefont {A.}~\bibnamefont {Mizutani}}, \
  and\ \bibinfo {author} {\bibfnamefont {H.-K.}\ \bibnamefont {Lo}},\ }\href
  {\doibase 10.1038/ncomms13523} {\bibfield  {journal} {\bibinfo  {journal}
  {Nature Communications 2016 7:null}\ }\textbf {\bibinfo {volume} {7}},\
  \bibinfo {pages} {ncomms13523} (\bibinfo {year} {2016})}\BibitemShut
  {NoStop}%
\bibitem [{\citenamefont {Azuma}\ and\ \citenamefont
  {Kato}(2016)}]{2016.ArXiv.Azuma.AggQuantRep}%
  \BibitemOpen
  \bibfield  {author} {\bibinfo {author} {\bibfnamefont {K.}~\bibnamefont
  {Azuma}}\ and\ \bibinfo {author} {\bibfnamefont {G.}~\bibnamefont {Kato}},\
  }\href {http://arxiv.org/abs/1606.00135} {\  (\bibinfo {year} {2016})},\
  \Eprint {http://arxiv.org/abs/1606.00135} {arXiv:1606.00135} \BibitemShut
  {NoStop}%
\bibitem [{\citenamefont {Schoute}\ \emph {et~al.}(2016)\citenamefont
  {Schoute}, \citenamefont {Mancinska}, \citenamefont {Islam}, \citenamefont
  {Kerenidis},\ and\ \citenamefont
  {Wehner}}]{2016.ArXiv.Schoute-Wehner.PerfQnetroute}%
  \BibitemOpen
  \bibfield  {author} {\bibinfo {author} {\bibfnamefont {E.}~\bibnamefont
  {Schoute}}, \bibinfo {author} {\bibfnamefont {L.}~\bibnamefont {Mancinska}},
  \bibinfo {author} {\bibfnamefont {T.}~\bibnamefont {Islam}}, \bibinfo
  {author} {\bibfnamefont {I.}~\bibnamefont {Kerenidis}}, \ and\ \bibinfo
  {author} {\bibfnamefont {S.}~\bibnamefont {Wehner}},\ }\href
  {http://arxiv.org/abs/1610.05238} {\bibfield  {journal} {\bibinfo  {journal}
  {arXiv preprint}\ ,\ \bibinfo {pages} {1}} (\bibinfo {year} {2016})},\
  \Eprint {http://arxiv.org/abs/1610.05238} {arXiv:1610.05238} \BibitemShut
  {NoStop}%
\bibitem [{\citenamefont {Ac{\'{i}}n}\ \emph {et~al.}(2007)\citenamefont
  {Ac{\'{i}}n}, \citenamefont {Cirac},\ and\ \citenamefont
  {Lewenstein}}]{2007.NaturePhys.Acin-Lewenstein.EntgPerc}%
  \BibitemOpen
  \bibfield  {author} {\bibinfo {author} {\bibfnamefont {A.}~\bibnamefont
  {Ac{\'{i}}n}}, \bibinfo {author} {\bibfnamefont {J.~I.}\ \bibnamefont
  {Cirac}}, \ and\ \bibinfo {author} {\bibfnamefont {M.}~\bibnamefont
  {Lewenstein}},\ }\href {\doibase 10.1038/nphys549} {\bibfield  {journal}
  {\bibinfo  {journal} {Nature Physics}\ }\textbf {\bibinfo {volume} {3}},\
  \bibinfo {pages} {256} (\bibinfo {year} {2007})},\ \Eprint
  {http://arxiv.org/abs/0612167} {arXiv:0612167 [quant-ph]} \BibitemShut
  {NoStop}%
\bibitem [{\citenamefont {{Van
  Meter}}(2014)}]{2014.Book.VanMeter.QuantumNetworking}%
  \BibitemOpen
  \bibfield  {author} {\bibinfo {author} {\bibfnamefont {R.}~\bibnamefont {{Van
  Meter}}},\ }\href
  {https://books.google.com/books?hl=en{\&}lr={\&}id=khmNAwAAQBAJ{\&}oi=fnd{\&}pg=PT9{\&}dq=info:uXh{\_}G0svvCkJ:scholar.google.com{\&}ots=aXYIo8gWtn{\&}sig=2WDF2pKUXrknmmbmqj6n9SCzC9M{\#}v=onepage{\&}q{\&}f=false}
  {\emph {\bibinfo {title} {{Quantum Networking.}}}}\ (\bibinfo  {publisher}
  {Wiley},\ \bibinfo {year} {2014})\BibitemShut {NoStop}%
\bibitem [{\citenamefont {Hayashi}\ \emph {et~al.}(2007)\citenamefont
  {Hayashi}, \citenamefont {Iwama}, \citenamefont {Nishimura}, \citenamefont
  {Raymond},\ and\ \citenamefont
  {Yamashita}}]{2007.STACS.Hayashi-Yamashita.QNetCoding}%
  \BibitemOpen
  \bibfield  {author} {\bibinfo {author} {\bibfnamefont {M.}~\bibnamefont
  {Hayashi}}, \bibinfo {author} {\bibfnamefont {K.}~\bibnamefont {Iwama}},
  \bibinfo {author} {\bibfnamefont {H.}~\bibnamefont {Nishimura}}, \bibinfo
  {author} {\bibfnamefont {R.}~\bibnamefont {Raymond}}, \ and\ \bibinfo
  {author} {\bibfnamefont {S.}~\bibnamefont {Yamashita}},\ }in\ \href {\doibase
  10.1007/978-3-540-70918-3_52} {\emph {\bibinfo {booktitle} {STACS 2007}}}\
  (\bibinfo  {publisher} {Springer Berlin Heidelberg},\ \bibinfo {address}
  {Berlin, Heidelberg},\ \bibinfo {year} {2007})\ pp.\ \bibinfo {pages}
  {610--621}\BibitemShut {NoStop}%
\bibitem [{\citenamefont {Kobayashi}\ \emph {et~al.}(2009)\citenamefont
  {Kobayashi}, \citenamefont {{Le Gall}}, \citenamefont {Nishimura},\ and\
  \citenamefont
  {R{\"{o}}tteler}}]{2009.Springer.Kobayashi-Roetteler.QnetCodewithClassComm}%
  \BibitemOpen
  \bibfield  {author} {\bibinfo {author} {\bibfnamefont {H.}~\bibnamefont
  {Kobayashi}}, \bibinfo {author} {\bibfnamefont {F.}~\bibnamefont {{Le
  Gall}}}, \bibinfo {author} {\bibfnamefont {H.}~\bibnamefont {Nishimura}}, \
  and\ \bibinfo {author} {\bibfnamefont {M.}~\bibnamefont {R{\"{o}}tteler}},\
  }in\ \href {\doibase 10.1007/978-3-642-02927-1_52} {\emph {\bibinfo
  {booktitle} {Lecture Notes in Computer Science (including subseries Lecture
  Notes in Artificial Intelligence and Lecture Notes in Bioinformatics)}}},\
  Vol.\ \bibinfo {volume} {5555 LNCS}\ (\bibinfo {year} {2009})\ pp.\ \bibinfo
  {pages} {622--633},\ \Eprint {http://arxiv.org/abs/0908.1457}
  {arXiv:0908.1457} \BibitemShut {NoStop}%
\bibitem [{\citenamefont {Satoh}\ \emph {et~al.}(2012)\citenamefont {Satoh},
  \citenamefont {{Le Gall}},\ and\ \citenamefont
  {Imai}}]{2012.PRA.Satoh-Imai.QuantNetCodeQRep}%
  \BibitemOpen
  \bibfield  {author} {\bibinfo {author} {\bibfnamefont {T.}~\bibnamefont
  {Satoh}}, \bibinfo {author} {\bibfnamefont {F.}~\bibnamefont {{Le Gall}}}, \
  and\ \bibinfo {author} {\bibfnamefont {H.}~\bibnamefont {Imai}},\ }\href
  {\doibase 10.1103/PhysRevA.86.032331} {\bibfield  {journal} {\bibinfo
  {journal} {Physical Review A}\ }\textbf {\bibinfo {volume} {86}},\ \bibinfo
  {pages} {032331} (\bibinfo {year} {2012})}\BibitemShut {NoStop}%
\bibitem [{\citenamefont {Satoh}\ \emph {et~al.}(2016)\citenamefont {Satoh},
  \citenamefont {Ishizaki}, \citenamefont {Nagayama},\ and\ \citenamefont {{Van
  Meter}}}]{2016.PRA.Satoh-VanMeter.QNetCodeAn}%
  \BibitemOpen
  \bibfield  {author} {\bibinfo {author} {\bibfnamefont {T.}~\bibnamefont
  {Satoh}}, \bibinfo {author} {\bibfnamefont {K.}~\bibnamefont {Ishizaki}},
  \bibinfo {author} {\bibfnamefont {S.}~\bibnamefont {Nagayama}}, \ and\
  \bibinfo {author} {\bibfnamefont {R.}~\bibnamefont {{Van Meter}}},\ }\href
  {\doibase 10.1103/PhysRevA.93.032302} {\bibfield  {journal} {\bibinfo
  {journal} {Physical Review A}\ }\textbf {\bibinfo {volume} {93}},\ \bibinfo
  {pages} {032302} (\bibinfo {year} {2016})}\BibitemShut {NoStop}%
\bibitem [{\citenamefont {Lemr}\ \emph {et~al.}(2013)\citenamefont {Lemr},
  \citenamefont {Bartkiewicz}, \citenamefont {{\v{C}}ernoch},\ and\
  \citenamefont {Soubusta}}]{2013.PRA.Lemretal.LinOptRout}%
  \BibitemOpen
  \bibfield  {author} {\bibinfo {author} {\bibfnamefont {K.}~\bibnamefont
  {Lemr}}, \bibinfo {author} {\bibfnamefont {K.}~\bibnamefont {Bartkiewicz}},
  \bibinfo {author} {\bibfnamefont {A.}~\bibnamefont {{\v{C}}ernoch}}, \ and\
  \bibinfo {author} {\bibfnamefont {J.}~\bibnamefont {Soubusta}},\ }\href
  {\doibase 10.1103/PhysRevA.87.062333} {\bibfield  {journal} {\bibinfo
  {journal} {Physical Review A}\ }\textbf {\bibinfo {volume} {87}},\ \bibinfo
  {pages} {062333} (\bibinfo {year} {2013})}\BibitemShut {NoStop}%
\bibitem [{\citenamefont {Pirandola}\ \emph {et~al.}(2009)\citenamefont
  {Pirandola}, \citenamefont {Garc{\'{i}}a-Patr{\'{o}}n}, \citenamefont
  {Braunstein},\ and\ \citenamefont {Lloyd}}]{2009.PRL.Pirandola-Lloyd.SKCLB}%
  \BibitemOpen
  \bibfield  {author} {\bibinfo {author} {\bibfnamefont {S.}~\bibnamefont
  {Pirandola}}, \bibinfo {author} {\bibfnamefont {R.}~\bibnamefont
  {Garc{\'{i}}a-Patr{\'{o}}n}}, \bibinfo {author} {\bibfnamefont {S.~L.}\
  \bibnamefont {Braunstein}}, \ and\ \bibinfo {author} {\bibfnamefont
  {S.}~\bibnamefont {Lloyd}},\ }\href {\doibase 10.1103/PhysRevLett.102.050503}
  {\bibfield  {journal} {\bibinfo  {journal} {Physical review letters}\
  }\textbf {\bibinfo {volume} {102}},\ \bibinfo {pages} {050503} (\bibinfo
  {year} {2009})}\BibitemShut {NoStop}%
\bibitem [{\citenamefont {Wilde}\ \emph {et~al.}(2017)\citenamefont {Wilde},
  \citenamefont {Tomamichel},\ and\ \citenamefont
  {Berta}}]{2017.IEEETransInfTheo.Wilde-Berta.PricComQChanStrongConv}%
  \BibitemOpen
  \bibfield  {author} {\bibinfo {author} {\bibfnamefont {M.~M.}\ \bibnamefont
  {Wilde}}, \bibinfo {author} {\bibfnamefont {M.}~\bibnamefont {Tomamichel}}, \
  and\ \bibinfo {author} {\bibfnamefont {M.}~\bibnamefont {Berta}},\ }\href
  {\doibase 10.1109/TIT.2017.2648825} {\bibfield  {journal} {\bibinfo
  {journal} {IEEE Transactions on Information Theory}\ }\textbf {\bibinfo
  {volume} {63}},\ \bibinfo {pages} {1792} (\bibinfo {year} {2017})},\ \Eprint
  {http://arxiv.org/abs/1602.08898} {arXiv:1602.08898} \BibitemShut {NoStop}%
\end{thebibliography}%

\appendix

\section{Distance metric for the local routing rule using $L^1$ norm and recursion}\label{sec:recursive}

\begin{figure}
\includegraphics[width=\columnwidth]{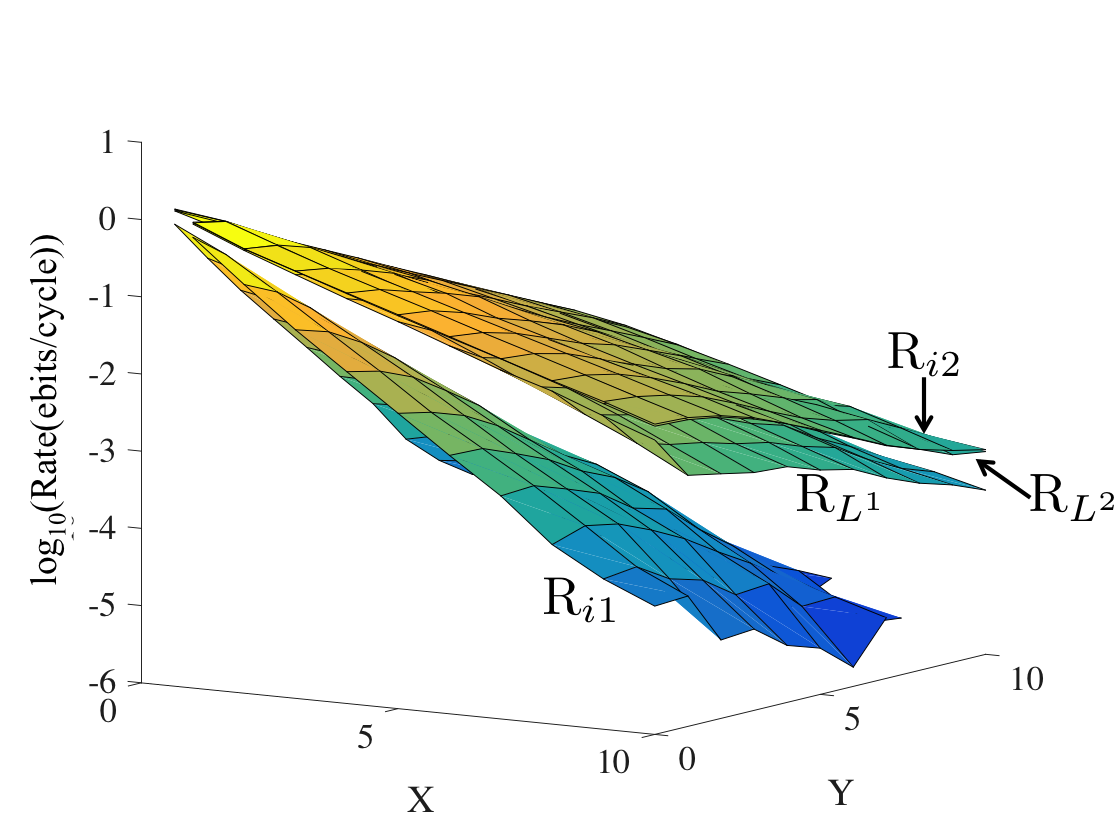}
\caption{Entanglement generation rates with different distance metrics. ${\textrm R}_{{L^1}}$ and ${\textrm R}_{{L^2}}$ are evaluated using the $L_1$ and $L_2$ norms respectively. The distance metric for ${\textrm R}_{{\textrm i1}}$ (iteration 1) is calculated using ${\textrm R}_{{L^1}}$, and ${\textrm R}_{{\textrm i2}}$ (iteration 2) is calculated using ${\textrm R}_{{\textrm i1}}$. ${\textrm R}_{{\textrm i2}}$ and ${\textrm R}_{{L^1}}$ are nearly indistinguishable as they coincide.}
\centering
\label{fig:recursive_rule}
\end{figure}

Our entanglement routing protocol with local link-state information uses the `distance' of neighboring repeater stations from Alice and Bob to decide which memories at a repeater should undergo entanglement swap attempts. The results presented in the paper use the $L^2$ norm as the distance metric. While the $L^2$ norm can be easily calculated for the square grid, it may not be easily generalizable for other (e.g., non-planar) topologies. Further, even though we do not prove the rate optimality of our local link-state routing protocol, given a network topology, it is not clear whether or not the $L^2$ norm is the optimal distance metric to be used in our protocol. 

In order to adapt our algorithm for arbitrary network topologies, and also to find a near-optimal distance metric for our algorithm, we employ the following numerical recursive method. Our evaluation begins with calculating ${\textrm R}_{{L^1}}({\textbf n_1}, {\textbf n_2})$, the entanglement generation rate achieved when our local rule is used to route entanglement between nodes ${\textbf n_1}$ and ${\textbf n_2}$, using the $L^1$ norm as the distance metric. In Fig.~\ref{fig:recursive_rule}, we plot ${\textrm R}_{{L^1}}({\textbf n_1}, {\textbf n_2})$ as a function of $(X,Y)$, where $X$ and $Y$ are the distance (in hops) between ${\textbf n_1}$ and ${\textbf n_2}$ along the horizontal and vertical dimensions of the square grid, respectively. The rate-distance scaling exponent for ${\textrm R}_{{L^1}}$ is worse than that of ${\textrm R}_{{L^2}}$, the rate attained by our protocol, using the $L^2$ norm as the distance metric. Next, for every repeater node ${\textbf n}$, we define distances $d_A$ and $d_B$ to Alice ${\textbf A}$ and Bob ${\textbf B}$ respectively, with respect to the following new distance metric (let us name this metric $i1$): $d_A := 1/{\textrm R}_{{L^1}}({\textbf n}, {\textbf A})$ and $d_B := 1/{\textrm R}_{{L^1}}({\textbf n}, {\textbf B})$. We then calculate ${\textrm R}_{{i1}}({\textbf n_1}, {\textbf n_2})$, the entanglement generation rate achieved when our local rule is used with the $i1$ distance metric to route entanglement between every pair of nodes ${\textbf n_1}$ and ${\textbf n_2}$. In Fig.~\ref{fig:recursive_rule}, we plot ${\textrm R}_{{i1}}({\textbf n_1}, {\textbf n_2})$ as a function of $(X,Y)$. We see that the rate-distance scaling achieved by ${\textrm R}_{{i1}}$ is even lower than that of ${\textrm R}_{{L^1}}$. However, when we go through the second iteration of the algorithm---i.e., define distance metric $i2$, under which $d_A = 1/{\textrm R}_{{i1}}({\textbf n}, {\textbf A})$ and $d_B = 1/{\textrm R}_{{i1}}({\textbf n}, {\textbf B})$, and use our local rule to evaluate ${\textrm R}_{{i2}}({\textbf n_1}, {\textbf n_2})$ as a function of $(X,Y)$---we find that the resulting rate ${\textrm R}_{{i2}}$ is almost the same (visually indistinguishable in the plot) as ${\textrm R}_{{L^2}}$, the rate we obtained directly when using the $L^2$ norm as the distance metric. This suggests that: (a) for the square grid (and presumably for any planar network topology) the $L^2$ norm metric might be near-optimal for use within our local rule, and that (b) for any given network topology, one could potentially pre-compute the optimal distance metric by a recursive strategy on the given topology using the $L^1$ norm as the starting point. However, there are instances where our local rule does not give the rate-optimal local routing rule. As an example, when $p = 1$ and $q = 1$, it is possible to find four disjoint paths without any link-state knowledge (the links are all deterministic) and the optimal rate is four ebits/cycle for any location of Alice and Bob. However, the fact that we are trying to route every flow through the best possible path without any coordination between different flows leads to collisions, which results in a rate that is below the optimal rate of four ebits/cycle.  Finding the rate-optimal local routing rule across different parameter values is left for future research.

%Whether or not the above recursive strategy attains the optimal distance metric for use with our local routing rule, and whether our local routing rule is the rate-optimal local routing rule, is being left open for future research.

\section{Multipath rate advantage}

\subsection{Numerical Evaluation}\label{sec:multipath_app}

\begin{figure}
\includegraphics[width=\columnwidth]{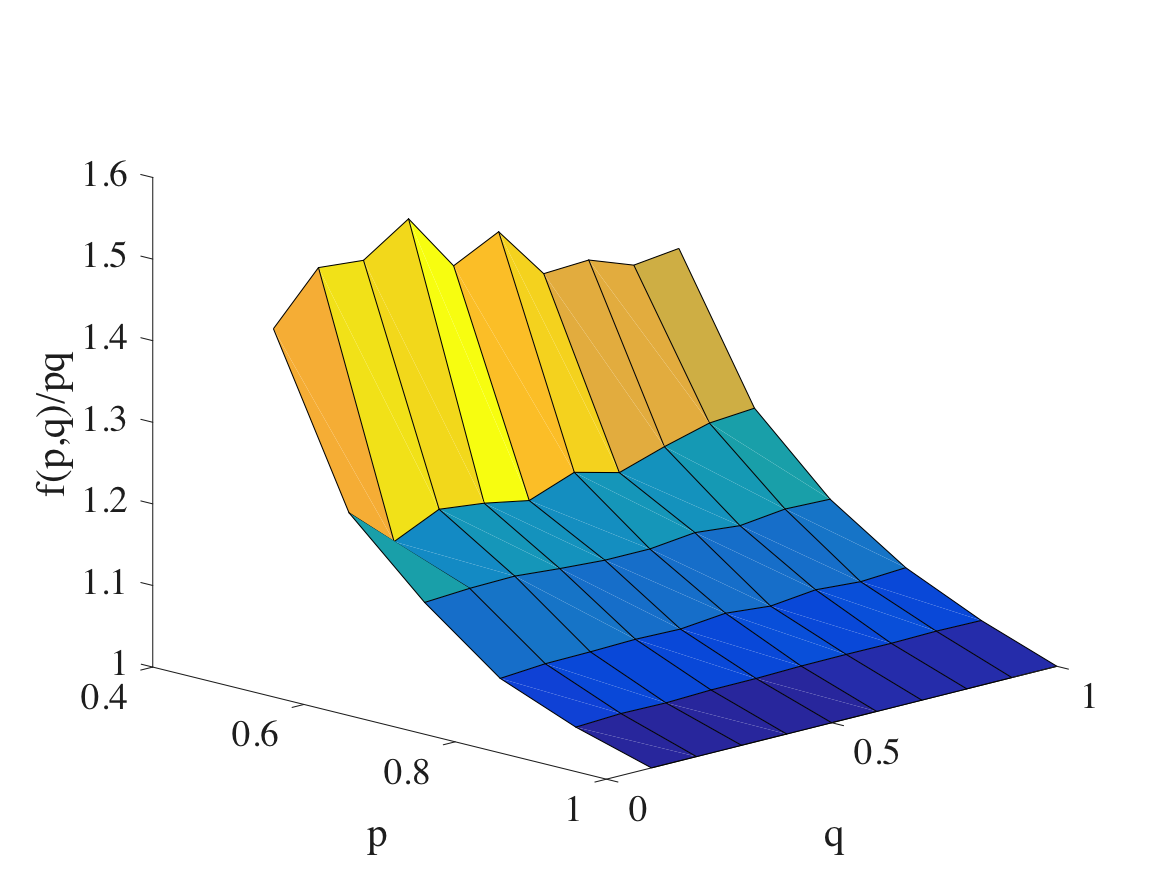}
\caption{$f(p,q)/pq$ quantifies the improvement in the scaling of $R_{\rm loc}(p,q)$ with respect to $R_{\rm lin}(p,q)$ with respect to the Alice-Bob Manhattan distance, $n$. $f(p,q)/pq$ increases as $p$ is reduced in $[1, p_c]$ but changing $q$ has a negligible effect.}
\centering
\label{fig:multipath_advantage}
\end{figure}

The goal this subsection is to quantify the improvement in the rate-vs.-distance exponent achieved by our local rule over that of a linear chain along the shortest path, for all possible pairs of values of $p$ and $q$. Fig.~\ref{fig: Rateplots}(c) shows this improvement, i.e., that of $R_{\rm loc}(p,q)$ compared to $R_{\rm lin}(p,q)$, for $p = 0.6$ and $q = 0.9$. Clearly, $R_{\rm lin}(p,q) = (pq)^{n}(p)/q \sim (1/q) [pq]^n$, where $n$ is the Manhattan distance between Alice and Bob. We have numerically verified that $R_{\rm loc}(p,q) \sim g(p,q)[f(p,q)]^n$ for $n$ large. We hence quantify the rate improvement by numerically evaluating the ratio $f(p,q)/(pq)$ exhaustively for all $(p, q) \in [0, 1] \times [0, 1]$, using Monte Carlo simulations. The results are shown in Fig.~\ref{fig:multipath_advantage}, for configurations of Alice and Bob located along $45\degree$ with respect to the grid axes. We see that $f(p,q)/pq$ increases as $p$ decreases in $[p_c, 1]$, but changing $q$ has a negligible effect on this ratio. 

%The intuitive reason for this behavior is that when $p > p_c$ (i.e., when the bond percolation criterion is met), the value of $p$ determines the expected number of edge disjoint paths between Alice and Bob in the percolated network instance after the first stage of each time step, but since connected path(s) are guaranteed via percolation, the end-to-end rate does not get a $p^n$ scaling. On the other hand, for a linear repeater chain to successfully produce a shared ebit between Alice and Bob, all links in the chain must be successful, the probability of which diminishes exponentially with the Alice-Bob shortest-path length (see discussion above). For multipath routing in the $p > p_c$ regime however, $q$ raised to the length of each edge-disjoint path is that path's expected rate contribution to $R_{\rm loc}$, which is analogous to $q^{{n_{\rm{SP}}(p)}-1}$ being an overall multiplier to the rate $R_{\rm lin}$. Since the total number of edge-disjoint paths can be at most the minimum of the degrees of Alice and Bob ($4$ in case of the square grid), the effect of $q$ in the rate-distance scaling for both $R_{\rm loc}$ and $R_{\rm lin}$ is similar, whereas $R_{\rm lin}$ suffers from an additional $p^{n_{\rm{SP}}(p)}$ rate decay.

\subsection{Analytical lower bound on the rate achieved by the local routing rule}\label{sec:localrateLB}

\begin{figure}
\includegraphics[width=\columnwidth]{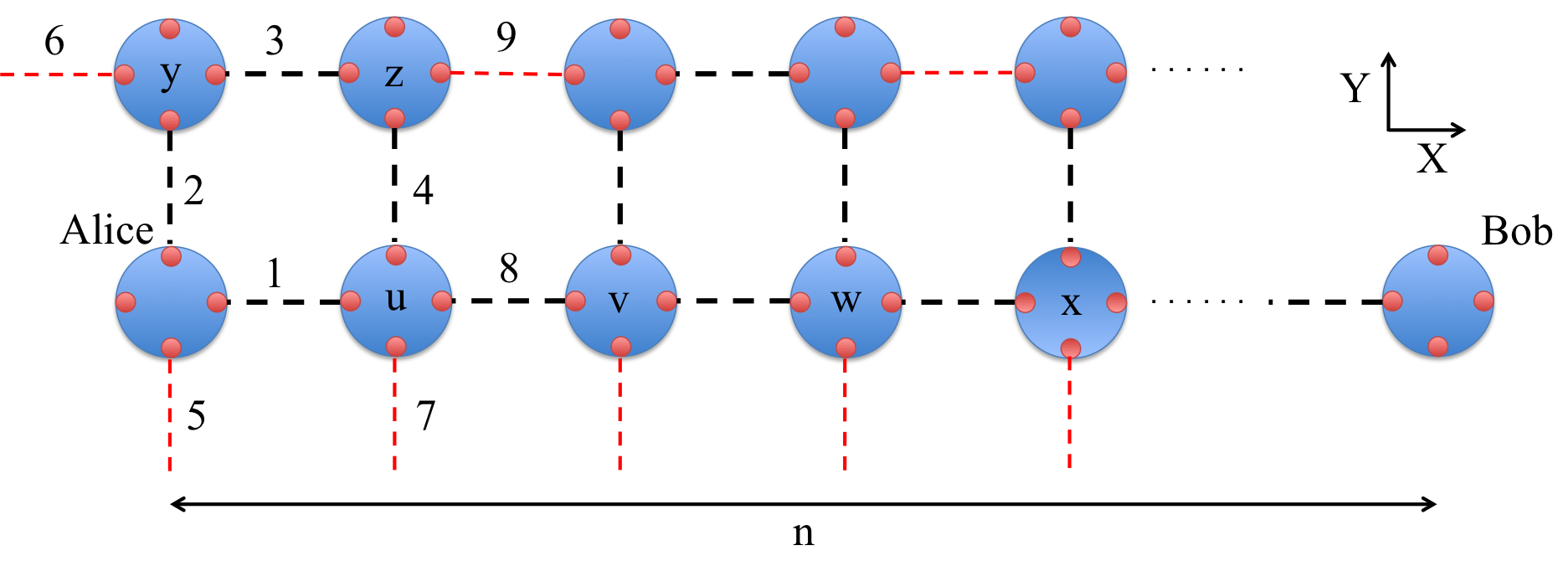}
\caption{Network used to prove the lower bound on entanglement generation rate with our local routing rule which shows that scaling of the rate with Alice-Bob manhattan distance for our rule is better than the scaling of the rate along a linear repeater chain along the shortest path between Alice-Bob.}
\centering
\label{fig:localLB_network}
\end{figure}

In this subsection, we derive an analytical lower bound on the entanglement generation rate attained by our local routing rule (using the $L^2$ norm as the distance metric), with the objective of demonstrating multi-path routing advantage, i.e., the rate-vs.-distance scaling attained by our local rule is strictly better than that attained by a linear repeater chain along the shortest path between Alice and Bob.

Consider routing entanglement between Alice and Bob located at $(X, Y)$ and $(X + n, Y)$ respectively, i.e., $n$ hops apart along the $X$ dimension of the square lattice. We will evaluate a lower bound on $R_{\rm loc}$ by only evaluating the rate contributions from paths in which all the (external) links belong to the set of black dashed links shown in Fig.~\ref{fig:localLB_network}. The choice of internal links made at repeater nodes proceed as usual per our local rule. As a result, there are instances in which our local rule routes entanglement through paths comprising not just the black links, resulting in flows that do not contribute to our rate lower bound. %The red dashed edges are used in our analysis but we don't count the contribution of flows that use these links, which suffices for our lower bound. The rate achieved using this restricted set of paths provides a lower bound on the actual rate attained over any set of paths. 

We will refer to Fig.~\ref{fig:localLB_network} for the ensuing discussion. Recall that external links succeed (are `up') with probability $p$ and fail (are `down') with probability $1-p$, whereas internal links succeed with probability $q$. Consider $P({A \leftrightarrow v})$, the probability that there is a path between Alice $A$ and repeater $v$ that uses only black links. $P({A \leftrightarrow v})$ includes the probability of making the required internal links to create a path between $A$ and $v$, but not the probability of any internal links at the end points $A$ or $v$. It is easy to see that in any given time step, there can be no more than one edge-disjoint path between $A$ and $v$ along the black dashed links, since link $8$ must be part of the path. Let $l$ $\left(\widetilde{l} \right)$ be the event that the external link $l$ is up (down). Further, note that at any given time step, of all the possible ($0$, $1$ or $2$) internal links attempted by our local rule at a repeater node, only one internal link, if successful, contributes to ${A \leftrightarrow v}$. Let $l-m$ be the event that the internal link attempted at a repeater node to connect external links $l$ and $m$ is successful. If links $1$ and $8$ are both up, node $u$ attempts to connect those two links based on our local rule, regardless of the other links. If links $2$, $3$, $4$ and $8$ are up, but $1$, $5$, $6$, $7$ and $9$ are down, $u$ attempts to connect $4$ and $8$, $z$ attempts to connect $3$ and $4$ and $y$ attempts to connect $2$ and $3$. Considering these two possibilities, we have

\begin{eqnarray}
%P({A \leftrightarrow v}) &>& \Big[ \textrm{Pr}(1) + \textrm{Pr}(2, 3, 4, \widetilde{1}, \widetilde{5}, \widetilde{6}, \widetilde{7}, \widetilde{9}, y, z)\Big] \textrm{Pr}(u,8) \nonumber \\
P({A \leftrightarrow v}) &>& \textrm{Pr}(1, 8, 1{\mymathhyphen}8) \nonumber \\
&& + \, \textrm{Pr}(2, 3, 4, 8, \widetilde{1}, \widetilde{5}, \widetilde{6}, \widetilde{7}, \widetilde{9}, 2{\mymathhyphen}3, 3{\mymathhyphen}4, 4{\mymathhyphen}8) \nonumber \\
&=& \left[ p + p^3(1-p)^5q^2 \right] pq \nonumber \\
&=& p'pq, \label{PAv}
\end{eqnarray}

\noindent where $p' = p + p^3(1-p)^5q^2 > p$. %If all four external links neighboring node $u$ succeed, only the internal link connecting $1$ and $8$ contributes to the entanglement generation rate. In this case, Pr($u$) is the probability that this internal link succeeded (the success of the link between $4$ and $7$ is irrelevant).

$P(v \leftrightarrow x)$ is the probability that there is a path between $v$ and $x$ that uses only black links (the probability of internal link successes at the end points $v$ and $x$ are not included). $P(v \leftrightarrow x)$ and $P({A \leftrightarrow v})$ are not independent events because they both involve link $9$. $P(v \leftrightarrow x | A \leftrightarrow v)$ is the probability that there exists a path along black dashed lines between $v$ and $X$ given that a path along black dashed lines exists between $A$ and $v$. We now show that $P(v \leftrightarrow x | A \leftrightarrow v) > P(v \leftrightarrow x)$.

\begin{eqnarray}
P (v \leftrightarrow x | A \leftrightarrow v) &=& P (v \leftrightarrow x | A \leftrightarrow v, 9)\textrm{Pr}(9| A \leftrightarrow v) + \nonumber \\
& & P \left(v \leftrightarrow x | A \leftrightarrow v, \widetilde{9}\right)\textrm{Pr}\left(\widetilde{9}| A \leftrightarrow v\right) \nonumber \\
&=& P (v \leftrightarrow x | 9)\textrm{Pr}(9| A \leftrightarrow v) + \nonumber \\
& & P \left(v \leftrightarrow x | \widetilde{9}\right)\textrm{Pr}\left(\widetilde{9} | A \leftrightarrow v\right) \nonumber \\
&=& P (v \leftrightarrow x | 9) \left(1-\textrm{Pr} \left(\widetilde{9} | A \leftrightarrow v \right)\right) + \nonumber \\
& & P \left(v \leftrightarrow x | \widetilde{9}\right)\textrm{Pr}\left(\widetilde{9} | A \leftrightarrow v \right) \nonumber \\
&=& \textrm{Pr}\left(\widetilde{9}| A \leftrightarrow v \right) \times \nonumber \\ 
& & \left(P \left(v \leftrightarrow x | \widetilde{9}\right) - P (v \leftrightarrow x | 9)\right) \nonumber \\
& & + P (v \leftrightarrow x | 9) \label{Pvxconditional}
\end{eqnarray}

\noindent where $P (v \leftrightarrow x | A \leftrightarrow v, 9) = P (v \leftrightarrow x | 9)$ and $P \left(v \leftrightarrow x | A \leftrightarrow v, \widetilde{9}\right) = P \left(v \leftrightarrow x |, \widetilde{9}\right)$ because link $9$ being up or down is the only probabilistic event that influences both $P(A \leftrightarrow v)$ and $P(v \leftrightarrow x)$. Further,

\begin{eqnarray}
P (v \leftrightarrow x) &=& P (v \leftrightarrow x | 9)\textrm{Pr}(9) + P \left(v \leftrightarrow x | \widetilde{9}\right)\textrm{Pr}\left(\widetilde{9}\right) \nonumber \\
&=& P (v \leftrightarrow x | 9) \left(1-\textrm{Pr} \left(\widetilde{9} \right)\right) + \nonumber \\
& & P \left(v \leftrightarrow x | \widetilde{9}\right)\textrm{Pr}\left(\widetilde{9}\right) \nonumber \\
&=& \textrm{Pr}\left(\widetilde{9}\right) \left(P \left(v \leftrightarrow x | \widetilde{9}\right) - P (v \leftrightarrow x | 9)\right) \nonumber \\
& & + P (v \leftrightarrow x | 9). \label{Pvx}
\end{eqnarray}

Comparing \ref{Pvxconditional} and \ref{Pvx}, $\textrm{Pr}\left(\widetilde{9}| A \leftrightarrow v \right)  = \textrm{Pr}\left(\widetilde{9}\right)\textrm{Pr}\left(A \leftrightarrow v |\widetilde{9}\right)/\textrm{Pr}\left(A \leftrightarrow v \right) >  \textrm{Pr}\left(\widetilde{9}\right)$ because $\textrm{Pr}\left(A \leftrightarrow v |\widetilde{9}\right) > \textrm{Pr}\left(A \leftrightarrow v \right)$ following equation \ref{PAv}. Similarly, $\left(P \left(v \leftrightarrow x | \widetilde{9}\right) - P (v \leftrightarrow x | 9)\right) > 0 $. Hence, $P(v \leftrightarrow x | A \leftrightarrow v) > P(v \leftrightarrow x)$.

From Fig.~\ref{fig:localLB_network}, we can see that in order to get a path along black dashed lines from $A$ to $x$, there must be a path along black dashed lines from $A$ to $v$ and from $v$ to $x$, and the internal link at $v$ must succeed. Therefore, 

\begin{eqnarray}
P (A \leftrightarrow x) &=& P (A \leftrightarrow v)qP (v \leftrightarrow x | A \leftrightarrow v) \nonumber \\
&>& P (A \leftrightarrow v)P (v \leftrightarrow x)q \nonumber \\
&=& \left( P (A \leftrightarrow v) \right)^2q \nonumber \\
&=& (p'pq)^2q,
\end{eqnarray}

%Following a reasoning similiar to $P({A \leftrightarrow v})$, the probability of a path between $v$ and $x$, inclusive of the success probability of internal links at $v$ and $x$, is $P_{v \leftrightarrow x} > p'pq^2$ where the additional factor of $q$ comes from the additional internal link at $v$, compared to $P_{A \leftrightarrow v}$. 

\noindent where we use symmetry between $A \leftrightarrow v$ and $v \leftrightarrow x$ in the third line. Repeating this for all repeaters between Alice and Bob, it is easy to see that

\begin{eqnarray}
R_{\rm loc} > P(A \leftrightarrow B) &>& p'^{\lceil n/2 \rceil} p^{\lfloor n/2 \rfloor}q^{n-1} \\
&\geq& \left(\sqrt{p'p}\right)^n q^{n-1} \nonumber \\
&=&  \left[\left(\sqrt{p'p}\right)q\right]^n q^{-1} \nonumber \\
&=& \left(pq\right)^{\beta n} q^{-1}, \nonumber
\end{eqnarray}

\noindent where $\lceil n/2 \rceil$ is the smallest integer greater than or equal to $n/2$ and $\lfloor n/2 \rfloor$ is the largest integer smaller than or equal to $n/2$. The second inequality uses the fact that $p' > p$ and $n > 0$. $\beta = \log\left[ \left( \sqrt{p'p} \right) q \right]/\log\left[pq \right] < 1$ because $p<p'<1$ and $q<1$.

Therefore, since $R_{\rm loc} > \left(pq\right)^{\beta n} q^{-1}$ with $\beta<1$ and $R_{\rm lin} = \left(pq\right)^{n} q^{-1}$, the exponent in the scaling with $n$ is smaller in $R_{\rm loc}$ compared to $R_{\rm lin}$, i.e. the rate-vs.-distance scaling is better with multi-path routing. Using a similar reasoning, it is easy to see that the same is true even when Alice and Bob are at located at different $Y$ coordinates. It should be noted that the lower bound we derive here is not meant to be tight (see Section~\ref{sec:multipath_app} for a full numerical evaluation of the exponents for $R_{\rm loc}$ and $R_{\rm lin}$). The only purpose of this subsection was to prove that the rate-vs.-distance scaling for entanglement routing strictly benefits from multi-path routing.
\end{document}